%% file: DFSpaper.tex
\documentclass[twocolumn,showpacs,preprintnumbers,amsmath,amssymb,pra]{revtex4}

\usepackage{graphicx}

\newcommand{\e}{\mathrm{e}}
\newcommand{\ii}{\mathrm{i}}

\newcommand{\Rzm}{\gate{R^z(-\frac{\pi}{2})}}
\newcommand{\Rzp}{\gate{R^z(\frac{\pi}{2})}}
\newcommand{\Rxm}{\gate{R^x(-\frac{\pi}{2})}}
\newcommand{\Rxp}{\gate{R^x(\frac{\pi}{2})}}
\newcommand{\RzmL}{\gate{R_L^z(-\frac{\pi}{2})}}
\newcommand{\RzpL}{\gate{R_L^z(\frac{\pi}{2})}}
\newcommand{\RxmL}{\gate{R_L^x(-\frac{\pi}{2})}}
\newcommand{\RxpL}{\gate{R_L^x(\frac{\pi}{2})}}

\newcommand{\eh}{\mathrm{e}}
\newcommand{\LL}{L}
\newcommand{\RR}{\mathrm{R}}

\input{Qcircuit}

\begin{document}

\title{Robust implementations of quantum repeaters}

\author{Alexander Klein, Uwe Dorner, Carolina Moura Alves, and Dieter Jaksch}

\affiliation{Clarendon Laboratory, University of Oxford, Parks Road,
Oxford OX1 3PU, United Kingdom}

\begin{abstract}
  We show how to efficiently exploit decoherence free subspaces (DFSs),
which are immune to collective noise, for realizing quantum
repeaters with long lived quantum memories. Our setup consists of an
assembly of simple modules and we show how to implement them in
systems of cold, neutral atoms in arrays of dipole traps. We develop
methods for realizing robust gate operations on qubits encoded in a
DFS using collisional interactions between the atoms. We also give a
detailed analysis of the performance and stability of all required
gate operations and emphasize that all modules can be realized with
current or near future experimental technology.
\end{abstract}

\date{\today}

\maketitle

\section{Introduction}

One of the major obstacles in the realization of quantum information
processors (QIP) is decoherence, caused by the unwanted interaction
of the system with its environment. Even the feasibility of simple
special purpose quantum devices, such as quantum repeaters, is
significantly affected by the presence of decoherence and a special
precaution to counteract the resulting loss of quantum information
has to be taken. A number of concepts have been developed to fight
decoherence and increase the reliability of QIP including active
error correction~\cite{Shor95,Steane96,Ekert96} and fault tolerant
quantum computing~\cite{Knill98}. Aside from this it was shown that,
dependent on the type of interaction processes between the qubits
and their environment, subspaces of the system's Hilbert space can
exist which are immune to decoherence processes induced by this
interaction~\cite{Zanardi97,Duan98,Lidar98,Bacon00}.
These decoherence free subspaces are a method of passive error
correction or error prevention and can significantly increase the
lifetime of quantum information and reliability of QIP as already
demonstrated with ion traps, NMR and optical
methods~\cite{Kielpinski01,Langer05,Fortunato02,Ollerenshaw03,Viola01,Kwiat00,Mohseni03,Bourennane04}.
However, this robustness makes states within the decoherence free
subspace (DFS) difficult to manipulate via controlled operations
and, in general, complicated interactions with a DFS are necessary
for realizing universal QIP~\cite{Bacon00}.

A particularly promising application of DFSs are quantum
repeaters~\cite{Briegel98,Duan01} which require very long quantum
information storage times but only a limited set of quantum
manipulations. In this paper we show that the quantum operations
necessary for a quantum repeater can be realized on DFS qubits with
nearest neighbor hopping~\cite{Pachos03,Calarco-PRA-00} and
interactions with ancilla qubits~\cite{Calarco04} only, thus
circumventing the difficulties usually associated with the usage of
DFS~\cite{Bacon00}. Quantum repeaters are used to distribute
maximally entangled pairs of qubits over long distances which is a
necessary prerequisite for important applications such as quantum
cryptography~\cite{Ekert91} and quantum
teleportation~\cite{Bennett93}. They are thus considered one of the
most important near future special purpose QIP.

The basic idea of a quantum repeater is to divide the transmission
line into segments with a length of the order of the attenuation
length of the channel. On each segment entangled particle pairs are
created and by applying entanglement swapping~\cite{Zukowski93} and
purification protocols~\cite{Bennett96a,Deutsch96,Briegel98}
entangled pairs of larger distances are produced. By repeating these
steps a distant pair of qubits with high entanglement fidelity is
obtained. Entanglement purification, distillation, and concentration
have been realized in a number of seminal
experiments~\cite{Pan03,Yamamoto03,Kwiat01,Zhao03} using photonic
states for encoding the qubits. Techniques for converting flying
photonic qubits into atomic ones via cavity-enhanced
interactions~\cite{Cirac97,vanEnk-PRL-1997} or the usage of atomic
ensembles~\cite{Fleischhauer-PRL-2000,Philipps-PRL-2001} have been
proposed. Very recently a demonstration of long distance
entanglement of massive particles has been
achieved~\cite{Kuzmich05}.

Here we develop a proposal for a quantum repeater with DFS quantum
memory using the entanglement purification scheme introduced
in~\cite{Briegel98,Duer99} (also known as the ``Innsbruck
protocol''). During the purification procedure the entanglement of a
primary pair of qubits is increased by sacrificing auxiliary
entangled qubit pairs. The entanglement fidelity of the final
entangled qubit pair does not converge to 1, but to a value which
depends on the fidelity of the initial primary pair and on the
constant fidelity of the auxiliary pairs.
However, the scheme~\cite{Briegel98,Duer99} is favorable from a
pragmatic point of view since it requires fewer physical resources
than the purification schemes used
in~\cite{Bennett96a,Bennett96b,Deutsch96}. On the other side, the
protocol~\cite{Briegel98,Duer99} takes longer and thus quantum
information has to be stored reliably for the whole duration of the
protocol. We overcome this apparent disadvantage by the usage of DFS
qubits of four neutral atoms which form the key elements of our DFS
quantum repeater.

Our scheme is based on current and near future technology and is
motivated by the extraordinary progress of the field of cold,
neutral atoms in dipole traps over the past few years both on the
theoretical~\cite{Jaksch98,Jaksch05,Cirac04,Santos-PRL-2004} and on
the experimental~\cite{Darquie05,Folling-Nature-2005} side. Atomic
quantum registers have been realized in arrays of
microlenses~\cite{Dumke02,Dumke02a} and by exploiting the superfluid
to Mott-Insulator transition in optical
lattices~\cite{Greiner02,stoferle-PRL-2004,Jaksch98}, single and
many atoms have been manipulated
coherently~\cite{Widera04,Mandel03}, atoms stored in standing waves
have been transported over macroscopic distances using conveyor belt
techniques~\cite{Nusse05,Sauer04,Kuhr03,Kuhr01}, and multipartite
entangled states of atoms were
created~\cite{Mandel03,Jaksch-PRL-99}. Furthermore, extremely
versatile technologies such as optical tweezers and holographic
traps that are controlled by liquid crystal spatial light
modulators~\cite{Grier03,Bergamini04,Dorner05} are currently
developed. They allow for essentially arbitrary designs of optical
potentials.

We will show that a DFS quantum repeater can be assembled from
modules which only require rotations of the logical DFS qubits
around the $x$ and $z$ axes by angles $\pi$ and $\pi/2$ and a
controlled phase (CPHASE) gate where an ancilla atom acts as the
control qubit. The single qubit rotations will be realized by
selectively lowering potential barriers between atoms and the system
will not leave the DFS during these gates. The ancilla qubit, which
is required to interact with a DFS qubit for realizing the CPHASE
gate, will not be decoherence free. However, we will show that its
influence on the DFS memory is sufficiently small to allow for gate
fidelities better than 98.7\%. Furthermore, we will show that our
scheme allows for deviations of the system parameters of a few
percent without significantly affecting the gate fidelities.

This paper is structured as follows: In Sec.~\ref{Sec:II} we will
briefly summarize the repeater protocol and show how DFSs can be
embedded into this scheme. The goal of this section is to show what
resources are needed for the implementation of the quantum repeater.
In Sec.~\ref{Sec:III} we propose methods of how the operations of
Sec.~\ref{Sec:II} can be physically realized with cold, neutral
atoms in arrays of dipole traps. The results are summarized in
Sec.~\ref{Sec:IV}.


\input{KapII}

\input{Implementation_new}


\section{Conclusion}
\label{Sec:IV}

In this paper we developed a scheme for the robust implementation of
a quantum repeater. The qubits at the repeater nodes are encoded in
a DFS which increases the decoherence time of the quantum memory
considerably and thus improves the reliability of the scheme. We
showed that the quantum repeater can be divided into simple modules
whose implementation only requires a small number of different
quantum gate operations. The single qubit gates can be realized by
selectively lowering potential barriers between the atoms without
ever leaving the DFS. Controlled operations are mediated by a single
auxiliary qubit which is not decoherence free. However, the
influence of this qubit on the gate fidelities was thoroughly
investigated and shown to be sufficiently small as not to
significantly affect the quality of the repeater scheme.

Furthermore, we presented proposals for implementing the DFS qubits
and all necessary operations in systems of neutral atoms trapped in
arrays of optical dipole traps. The times for implementing these
gates were calculated and shown to be much smaller than the expected
decoherence time of the DFS qubits. Numerical examinations have
shown that the achieved gate fidelities can be better than 98.7\%
and that the gate operations are stable against possible, small
deviations of the system parameters.

Our scheme is extendable to general purpose quantum computing with
DFS qubits, since the elementary one-qubit rotations $R^x_L$,
$R^z_L$ and a two-qubit gate, which together form a universal set of
gates, can be realized. However, during the CPHASE gate the DFS is
left. This effect is very small for our quantum repeater scheme and
the influence of leaving the DFS in other applications as well as
the scalability of the scheme is the subject of further
investigations.

Recent experiments have demonstrated that it is possible to
implement and control quantum registers with a few atoms in optical
lattices \cite{Schrader04} or dipole trap arrays \cite{Dumke02a} and
that the decoherence times of the auxiliary atom are long enough to
ensure a successful operation of our scheme \cite{Kuhr03, Kuhr05}.
The experimental combination of atom registers and optical conveyor
belts is also planned \cite{Schrader04}. Hence we are confident that
the presented scheme can be implemented with current and near future
experimental techniques.

\acknowledgments{A.K.~thanks Stephen R.~Clark for useful discussions
and acknowledges financial support from the Keble Association. This
work was supported by the EPSRC (UK) through the QIP IRC
(GR/S82176/01) and project EP/C51933/1. The research was also
supported by the EU through the STREP project OLAQUI. The research
of U.D.~was supported by a Marie Curie Intra-European Fellowships
within the 6th European Community Framework Programme (`RAQUIN').
The contents of this paper reflects the author's views and not the
views of the European Commission. C.M.A.~is supported by the
Funda\c{c}\~{a}o para a Ci\^{e}ncia e Tecnologia (Portugal). }

\appendix

\section{Adiabatic elimination for the $X_\LL$-gate \label{AppA}}
The derivation of Hamiltonian $H_X$, Eq.~(\ref{HamilX}), is given in
more detail. The starting point is the Hamiltonian
Eq.~(\ref{Hamiltonian}) for three atoms in three lattice sites.
Since the number of atoms in state $a$ and $b$ is conserved the
states decouple in two subspaces where the first one contains only
states with one atom in state $a$ and two atoms in state $b$ and the
other one contains two atoms in state $a$ and one atom in state $b$.
The states with three atoms in the same internal state do not occur
in our system. For simplicity we only consider the second subset,
the results for the other subset can be derived in the same way.

Let $Q$ be the projection operator on all states with more than one
atom in a lattice site, $P = \openone -Q$. The Hamiltonian after
adiabatic elimination in first order perturbation theory is given by
\begin{equation}
  \hat H_\mathrm{eff} = \lim_{\varepsilon \to 0} P\hat H P -
  P \hat H Q (Q \hat H Q + \varepsilon \openone)^{-1} Q \hat H P \,.
\end{equation}
The inversion of $Q \hat H Q + \varepsilon \openone$ was calculated
up to first order in the perturbation parameter $J/U$ since in this
treatment higher orders of $J/U$ are neglected anyway. In the basis
$\left\{ \ket{100},\ket{010},\ket{001} \right\}$ the adiabatically
eliminated Hamiltonian has the form
\begin{equation}
  \hat H_\mathrm{eff} = \left(\begin{array}{ccc}
  f_1 & g_1& 0 \\
  g_1& f_2 & g_2 \\
  0 & g_2 & f_3
  \end{array} \right)
\end{equation}
with
\begin{gather}
  f_1 = - \frac{U_a \left((J_1^{(a)})^2 + (J_1^{(b)})^2\right)
  + 4 U_{ab}(J_2^{(a)})^2 }
  {U_{a}U_{ab}} \,, \\
  f_2 = - \frac{(J_1^{(a)})^2+(J_2^{(a)})^2+(J_1^{(b)})^2+(J_2^{(b)})^2 }
  {U_{ab}} \,, \\
  f_3 = - \frac{U_a \left((J_2^{(a)})^2 + (J_2^{(b)})^2\right)
  + 4 U_{ab}(J_1^{(a)})^2 }
  {U_{a}U_{ab}} \,, 
\end{gather}
\begin{gather}
  g_1 = - 2 \frac{J_1^{(a)}J_1^{(b)}}{U_{ab}} \,, \\
  g_2 = - 2 \frac{J_2^{(a)}J_2^{(b)}}{U_{ab}} \,.
\end{gather}
The action of this Hamiltonian on the logical qubits can be written
in the form
\begin{gather}
  \hat H_\mathrm{eff} \ket{0}_\LL = u \ket{0}_\LL + v \ket{1}_\LL
  \,, \\
  \hat H_\mathrm{eff} \ket{1}_\LL = u \ket{1}_\LL + v \ket{0}_\LL
  \,,
\end{gather}
Solving the corresponding equations for $u$ and $v$ one finds
\begin{gather}
  u = f_2 \,,\\
  v = \frac{\sqrt{3}}{2} g_2 \,,
\end{gather}
\begin{gather}
  f_1 = f_2 + 2g_1 \,,\\
  g_2 = 2g_1 \,,\\
  f_3 = f_2 + g_1 \,.
\end{gather}
Because of the symmetry between the states with two atoms in state
$a$ and one in state $b$ and vice versa the only physical solution
for these equations is given by Eqs.~(\ref{Cond_X_J}) -
(\ref{HamilX}).

\section{Qubits remain in the DFS \label{AppB}}
In this appendix we show that the DFS qubits remain decoherence free
in the case without any interaction between the physical qubits,
which is relevant for the fast $R^z_L(\pi)$ rotation and the CPHASE
gate. Since in this case there can be more than one atom in a
lattice site we introduce Fock states
$\ket{n_a^1,n_b^1;n_a^2,n_b^2;...;n_a^5,n_b^5}$,  where $n_a^j$
($n_b^j$) is the number of atoms in mode $a$ ($b$) in lattice site
$j=1,...,5$ , respectively.

The fast $R^z_L(\pi)$ rotation can be performed by setting $U_{a,b}
= U_{ab}=0 $, $J^{(a,b)}_2=J^{(a,b)}_3=J^{(a,b)}_4=0$. In this case
the corresponding time evolution operator $U(t) = \exp(- \ii \hat H
t/\hbar)$, where $\hat H$ is given by Eq.~(\ref{Hamiltonian}), can
be calculated analytically and we get
\begin{equation}
  U(t) \ket{0}_L = \ket{0}_L \,,
\end{equation}
for all $t$ and thus the system never leaves the DFS. The state
$\ket{1}_L$ evolves as
\begin{widetext}
\begin{equation} \label{eq:B2}
\begin{split}
  U(t)& \ket{1}_L = \cos \left(\frac{2 J t}{\hbar} \right) \ket{1}_L
                  + \frac{\ii}{2 \sqrt{3}} \sin\left(\frac{2 J t}{\hbar} \right)
   \left\{  \frac{2}{\sqrt{2}} \left[ \ket{0,2;0,0} + \ket{0,0;0,2}
   \right] \otimes
   \ket{1,0;0,0;1,0} \right.  \\
   & +    \left.   \frac{2}{\sqrt{2}} \left[ \ket{2,0;0,0}
   + \ket{0,0;2,0} \right]\otimes \ket{0,1;0,0;0,1}
   - \left[\ket{1,1;0,0} + \ket{0,0;1,1} \right]
   \otimes \left[\ket{1,0;0,0;0,1} + \ket{0,1;0,0;1,0}
   \right]
   \right\} \,.
\end{split}
\end{equation}
Clearly, the first term $\cos(2Jt/\hbar) \ket{1}_L $ remains in the
DFS. In order to see that this is also true for the second part we
rewrite the states in first quantization. Let $\ket{\alpha \beta}_A$
denote a state where the first atom is in state $\ket{\alpha}$ and
the second one in state $\ket{\beta}$, independent in which lattice
site they are stored. Due to symmetrization we get
\begin{gather}
  \frac{1}{\sqrt{2}} \left[ \ket{1,1;0,0} + \ket{0,0;1,1} \right]
  \rightarrow \frac{1}{\sqrt{2}} \left[ \ket{01}_A + \ket{10}_A \right] \,,\\
  \frac{1}{\sqrt{2}} \left[ \ket{2,0;0,0}+ \ket{0,0;2,0}\right]
  \rightarrow \ket{00}_A \,,\\
  \frac{1}{\sqrt{2}} \left[ \ket{0,2;0,0}+ \ket{0,0;0,2}\right]
  \rightarrow \ket{11}_A \,.
\end{gather}
By rewriting Eq.~(\ref{eq:B2}) we get
\begin{equation}
\begin{split}
  U(t) \ket{1}_L = \cos \left(\frac{2 J t}{\hbar} \right) \ket{1}_L
                  + \frac{\ii}{2 \sqrt{3}} \sin\left(\frac{2 J t}{\hbar}
                  \right)
   &  \left\{  2 \ket{11}_A \otimes
   \ket{00}_A  + 2  \ket{00}_A \otimes
   \ket{11}_A  \right. \\
   & \qquad - \left.
   \left[\ket{01}_A + \ket{10}_A \right]
   \otimes \left[\ket{01}_A + \ket{10}_A
   \right]
   \right\} \,.
\end{split}
\end{equation}
The part proportional to the sine function is again decoherence
free, because it is just the state $\ket{1}_L$.

The case of the CPHASE gate is more complicated because the atoms
can hop into an additional lattice site. The time evolution of state
$\ket{0}_L$ is given by
\begin{equation}
\begin{split}
  U(t) \ket{0}_L =& \left[\ket{1,0;0,1} - \ket{0,1;1,0} \right]
  \otimes \left\{ \cos\left(
   \frac{\sqrt{2}J t}{\hbar} \right) \left[\ket{1,0;0,0;0,1} - \ket{0,1;0,0;1,0}
    \right] \right. \\
  & \left.\qquad+\frac{\ii}{\sqrt{2}} \sin \left( \frac{\sqrt{2} J t}{\hbar}\right)
 \left[  \ket{1,0;0,1;0,0} - \ket{0,1;1,0;0,0}
 +\ket{0,0;1,0;0,1} - \ket{0,0;0,1;1,0}  \right]
   \right\} \,.
\end{split}
\end{equation}
For $\ket{1}_L$ we get
\begin{equation}
\begin{split}
  U(t) \ket{1}_L =& \left[\ket{1,0;1,0}+ \ket{0,1;0,1} \right]\otimes \left\{ \frac{1}{4} \left(
   3 + \cos\left( \frac{2 \sqrt{2} J t}{\hbar} \right) \right)
   \left[ \ket{0,1;0,0;1,0} + \ket{1,0;0,0;0,1}  \right] \right.\\
   &\qquad+
   \frac{\ii}{2 \sqrt{2}} \sin\left(\frac{2 \sqrt{2} J t}{\hbar}
   \right) \left[ \ket{0,0;0,1;1,0} + \ket{0,0;1,0;0,1} +\ket{0,1;1,0;0,0} + \ket{1,0;0,1;0,0} \right]
    \\
   &\qquad- \left.\frac{1}{2} \sin^2\left( \frac{\sqrt{2} J t}{\hbar}\right)
   \left[  \ket{1,1;0,0;0,0}  + 2 \ket{0,0;1,1;0,0}+ \ket{0,0;0,0;1,1} \right]
   \right\}   \\
   &+
   \ket{0,1;0,1} \otimes \left\{
    \frac{1}{4} \left(
   3 + \cos\left( \frac{2 \sqrt{2} J t}{\hbar} \right) \right)
   \ket{1,0;0,0;1,0} \right. \\
   &\qquad+ \frac{\ii}{2 \sqrt{2}}\sin\left(\frac{2 \sqrt{2} J t}{\hbar}
   \right) \left[\ket{1,0;1,0;0,0} + \ket{0,0;1,0;1,0} \right] \\
   &\qquad-\left.\frac{1}{2\sqrt{2}} \sin^2\left( \frac{\sqrt{2} J t}{\hbar}\right)
   \left[  \ket{2,0;0,0;0,0}  + 2 \ket{0,0;2,0;0,0}+  \ket{0,0;0,0;2,0} \right]
   \right\}\\
   &+
   \ket{1,0;1,0} \otimes \left\{
    \frac{1}{4} \left(
   3 + \cos\left( \frac{2 \sqrt{2} J t}{\hbar} \right) \right)
   \ket{0,1;0,0;0,1} \right. \\
   &\qquad+ \frac{\ii}{2 \sqrt{2}}\sin\left(\frac{2 \sqrt{2} J t}{\hbar}
   \right) \left[\ket{0,1;0,1;0,0} + \ket{0,0;0,1;0,1} \right] \\
   &\qquad-\left.\frac{1}{2\sqrt{2}} \sin^2\left( \frac{\sqrt{2} J t}{\hbar}\right)
   \left[  \ket{0,2;0,0;0,0}  + 2 \ket{0,0;0,2;0,0}+  \ket{0,0;0,0;0,2} \right]
   \right\}
\end{split}
\end{equation}

Making analogous substitutions as above we see that these qubits are
again decoherence-free at all times.
\end{widetext}

\bibliography{DFSpaper}

\end{document}

%% file: KapII.tex
\section{Quantum Repeaters with decoherence free subspaces}
\label{Sec:II}

\begin{figure}[t!]
  \begin{center}
    \includegraphics[]{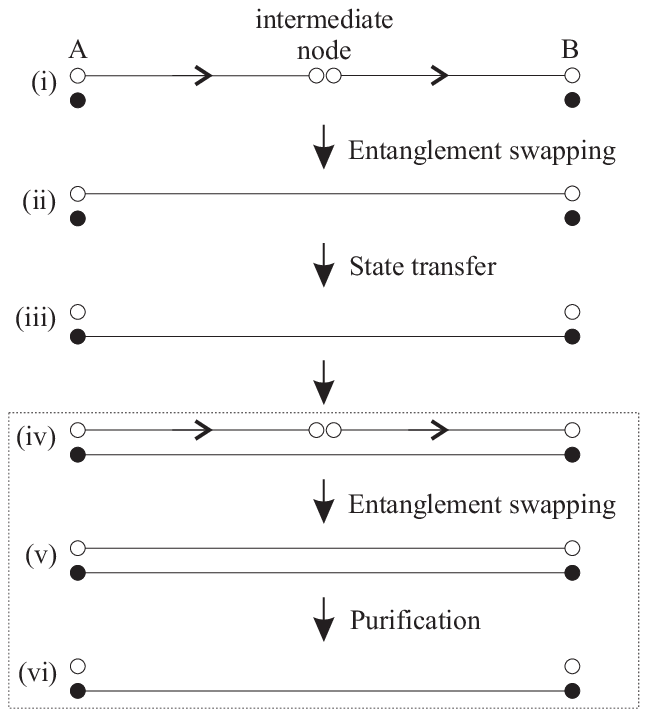}
    \caption{%
      Illustration of the entanglement purification scheme. Pairs of
      entangled flying qubits (open circles) are created via
      entanglement swapping.  These pairs are then used to purify an
      entangled pair of DFS qubits (solid circles; see text). The
      framed steps (iv)-(vi) are repeated until a desired entanglement
      fidelity is achieved.}
    \label{fig:repeater}
  \end{center}
\end{figure}

In the following we will describe the purification and repeater
protocol of Refs.~\cite{Briegel98,Duer99} and extend it by using
DFS. The basic steps of the purification scheme are illustrated in
Fig.~\ref{fig:repeater}. Only one quantum channel (represented by
the lines with an arrow) and two DFS qubits (represented by the
solid circles) are needed. The goal is to share a high fidelity Bell
state between $A$ and $B$ which is done by transmission of
``one-half'' of a maximally entangled flying qubit pair (open
circles in Fig.~\ref{fig:repeater}(i)). In general, the flying qubit
(e.g., a photon) is not directly sent from $A$ to $B$ since the
present available transmission quantum channels (particularly
optical fibers) degrade the entanglement of the flying qubit pair
exponentially with the distance between the nodes. Thus, if the
channel is too long, the minimum fidelity required for the
successful application of the purification protocol
($F_\mathrm{min}$) is not achieved.  To overcome this problem a
sufficiently high number of intermediate nodes between $A$ and $B$
is introduced such that the distance between two nodes is of the
order of the optical fiber's attenuation length~\cite{Duer99}.
Entangled flying qubit pairs are prepared at each of the
intermediate nodes and transmitted to the next node. In
Fig.~\ref{fig:repeater} this is illustrated for the example of one
intermediate node. After this, entanglement swapping
\cite{Zukowski93} at all intermediate nodes is performed which
yields a partially entangled pair of flying qubits between $A$ and
$B$ and its state is transferred to a DFS qubit pair (step (ii)).
Finally, as in (i) and (ii), auxiliary entangled qubit pairs are
generated which are sacrificed in order to increase the entanglement
of the DFS qubit pair (step (iv)-(vi)). Steps (iv)-(vi) are repeated
until a desired degree of entanglement is obtained for the DFS qubit
pair.

\begin{figure}[t!]
  \begin{center}
    \includegraphics[]{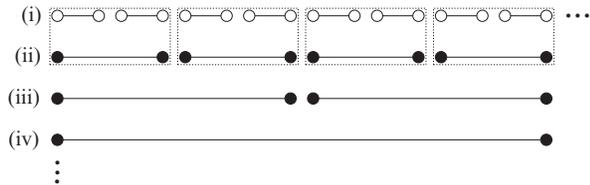}
    \caption{%
      The quantum repeater scheme. The dashed boxes indicate the
      subroutine shown in Fig.~\ref{fig:repeater}. The created
      entangled DFS qubit pairs are used to purify further DFS qubit
      pairs (in line (iii)). The resulting pairs are then used to
      purify the pairs of line (iv) and so on.}
    \label{fig:repeater1}
  \end{center}
\end{figure}

However, a straightforward extension of this scheme in order to
connect arbitrary separated points $A$ and $B$ is not possible since
the distance between the intermediate nodes has to be chosen such
that the fidelity of the final pair is not lower than
$F_\mathrm{min}$. This is, in general, not possible if the distance
between $A$ and $B$ is too large. Therefore, a quantum repeater
concept by means of a nested purification scheme was proposed
\cite{Briegel98,Duer99} which uses the above purification protocol
to create a number of entangled DFS qubit pairs which are then used
to purify a lower number of additional DFS pairs with greater
distance.  These pairs are then again used to purify an even lower
number of pairs, etc. until merely one entangled pair is left. This
procedure is illustrated in Fig.~\ref{fig:repeater1}: The boxes
stretching over lines (i) and (ii) represent the purification
protocol of Fig.~\ref{fig:repeater}. The entangled pairs of line
(ii) are then used to purify the pairs in line (iii) in the same way
as in Fig.~\ref{fig:repeater} except that the flying qubits have to
be replaced by DFS qubits, i.e., entanglement swapping between DFS
qubits, a state transfer between DFS qubits (to initialize the pairs
in line (iii) and (iv)) and purification by using entangled DFS
pairs have to be performed. The diagram shown in
Fig.~\ref{fig:repeater1} can be extended in horizontal and vertical
directions such that an arbitrary distance between the final pair
can be achieved. It was shown that the number of (DFS) qubits
necessary for the scheme increases merely logarithmically with the
distance of the final entangled pair~\cite{Briegel98,Duer99}.

In summary, the purification procedure shown in
Fig.~\ref{fig:repeater} can be realized by assembling the following
modules: (0) The creation of pairs of flying qubits and performing
entanglement swapping between them, (1) the realization of two DFS
qubits, (2) the transfer of the state of the flying qubits to the
DFS qubits, and (3) the actual purification of the entangled DFS
qubit pair.  Additionally, for the nested purification protocol,
i.e., for the full implementation of the quantum repeater, we have
to (4) perform entanglement swapping between neighboring DFS qubits,
(5) we need a state transfer between DFS qubits (analogously to
module (2)), and (6) we must perform entanglement purification using
two DFS qubit pairs. Module (0) can be implemented using standard
quantum optical methods as demonstrated
in~\cite{Pan98,Jennewein02,Sciarrino02,Takei05,Gisin05}. It does not
involve DFS qubits and will thus not be discussed further in this
paper.

In the following we will show how each of the remaining modules can
be realized. For the operations using ancilla qubits and thus
leaving the DFS we then calculate the achievable fidelities.

\subsection{Realization of the repeater modules \label{Sec:AllModules}}

\subsubsection{Module 1: Decoherence free memory and single qubit rotations}\label{SecII1}

The main sources of decoherence in systems of cold atoms
are fluctuations of the optical potential and external
electromagnetic (stray-)fields~\cite{Kuhr05}. These sources are
homogeneous across the region of the physical qubits
and the interaction with them thus only depends on their internal
states. The Hamiltonian describing the situation is given by
\begin{equation}
H_{I} = \sum_{i} \sigma^x_i B_x + \sigma^y_i B_y + \sigma^z_i B_z.
\label{decoherence_field}
\end{equation}
Here, $i$ labels the physical qubits, $\mathbf{B} = (B_x,B_y,B_z)$
describes the fields and fluctuations, and the $\sigma_i$ are Pauli
matrices acting on the $i$th physical qubit.

The corresponding decoherence free subspace is spanned by the states
with total zero angular momentum, and the lowest number of physical
qubits giving a twofold degenerate DFS is 4.  The DFS qubit is then
represented by~\cite{Bacon00,Kempe01}
\begin{equation}
\begin{split}
 \ket{0}_L &= \frac{1}{2}(\ket{01} - \ket{10})\otimes (\ket{01} - \ket{10}) \,,   \\
 \ket{1}_L &= \frac{1}{2\sqrt3} (2\ket{1100} + 2\ket{0011} - (\ket{01} + \ket{10})^{\otimes2}) \,,
\label{eq:logical_qubit}
\end{split}
\end{equation}
where $\ket{ijkl}=\ket{i}_1\ket{j}_2\ket{k}_3\ket{l}_4$ with
$i,j,k,l=0,1$ describes the computational basis states of the four
physical qubits. Each of the four physical qubits is represented by
two internal states of an atom which are trapped in an array of
dipole traps, for example, an optical lattice. By the method
described in Sec.~\ref{Sec:IIIa} the system is prepared in the
logical state $\ket{0}_L$ and in Sec.~\ref{Sec:IIIb} it is shown
that by tuning the system parameters appropriately we can implement
the Hamiltonians $H_X$ and $H_Z$ with
\begin{eqnarray}
 H_X &=& X_L = \frac{1}{\sqrt 3}(V_{12}+2V_{23}  ) \label{eq:HX}, \\
 H_Z &=& Z_L = -V_{12} \label{eq:HZ},
\end{eqnarray}
where $X_L$ ($Z_L$) is the $X$-gate ($Z$-gate) for the logical qubit
and $V_{ij}$ are permutations of the physical qubits $i$ and $j$. Note
that the logical Z-gate can also be represented as $Z_L = -V_{34}$.
Since the DFS defined by Eq.~(\ref{eq:logical_qubit}) is invariant
under permutations of the physical qubits~\cite{Bacon00}, the time
evolution of the system generates rotations
\begin{eqnarray}
 R_L^x(\theta) &=& \e^{-\ii \frac{\theta}{2} X_L }, \label{eq:rot_x}\\
 R_L^z(\theta) &=& \e^{-\ii \frac{\theta}{2} Z_L } \label{eq:rot_z}
\end{eqnarray}
{\em without ever leaving the DFS}, where $\theta=2t$ and $t$ is the
time necessary for performing the rotations. The index $L$ indicates
that these are operations on the logical qubit (in the following we
will also use this notation for operations on physical qubits, where
we omit the index $L$).

Although it is not required in the repeater protocol, we can thus
perform arbitrary rotations of the logical qubit (and hence we can
prepare arbitrary states in the DFS) since every two level unitary
operator can be decomposed in $R_L^{x}$ and $R_L^z$~\cite{Nielsen}.
\subsubsection{Module 2: State transfer between flying and DFS qubit}
\label{sec:R2}
\begin{figure}[h]
\[
\Qcircuit @C=1em @R=.7em {
  & \lstick{\ket{\psi}_q} & \qw    & \ctrl{1} & \gate{R^x(-\frac{\pi}{2})} & \meter & \control \cw \cwx[1]\\
  & \lstick{\ket{0}_L}    & \gate{R_L^x(\frac{\pi}{2})} & \gate{-Z_L} &
  \gate{R_L^x(-\frac{\pi}{2})} & \qw & \gate{R^z_L(\pi)} &
  \rstick{\ket{\psi}_L } \qw }
\]
\caption{%
  Quantum circuit for the state transfer of the auxiliary qubit to the
  DFS qubit.}
\label{fig:circuit1a}
\end{figure}
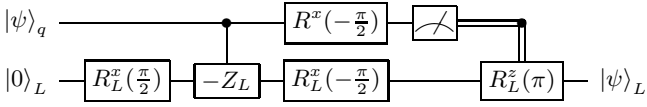

The quantum circuit for the state transfer of the flying qubit to the
logical qubit is shown in Fig.~\ref{fig:circuit1a}.
The state of the flying qubit is denoted as $\ket{\psi}_q =
\alpha\ket{0}_q + \beta\ket{1}_q$. As described in
Sec.~\ref{Sec:III} the state of the flying qubit (a photon) is first
transferred to an appropriate internal state of an auxiliary atom.
The state of this atom then controls the interaction between
physical qubits $3$ and $4$ leading to a controlled SWAP operation
between these atoms.  Since $Z_L=-V_{34}$, a controlled SWAP
operation of the physical qubits $3$ and $4$ leads to a controlled
$(-Z_L)$ (CPHASE) gate on the logical qubit.  As can be seen from
Fig.~\ref{fig:circuit1a}, apart from rotations given in
Eqs.~(\ref{eq:rot_x}) and~(\ref{eq:rot_z}), we need to perform a
rotation $R^x(-\pi/2)$ and a measurement of the control atom which
can be done by standard techniques such as laser induced Rabi
oscillations and state selective laser excitations.  Eventually, the
state of the atom is transferred to the logical qubit, i.e.,
$\ket{\psi}_L = \alpha\ket{0}_L + \beta\ket{1}_L$.

\subsubsection{Module 3: Entanglement purification}
\label{sec:R3}
The local operations required for entanglement purification are shown
in Fig.~\ref{fig:circuit_purif}.
\begin{widetext}
\begin{center}
\begin{figure}[h!]
\[
\Qcircuit @C=1em @R=.7em {
  \lstick{\text{auxiliary qubit}} & \gate{U^{A,B}}   & \Rzm & \Rxm               & \ctrl{1}    & \Rxp              & \Rzp & \meter \\
  \lstick{\text{DFS qubit}}       & \gate{U_L^{A,B}} & \qw  & \gate{R_L^x(-\pi)} & \gate{-Z_L} & \gate{R_L^x(\pi)} & \qw  & \qw }
\]
\caption{%
  Quantum circuit for entanglement purification. The procedure is
  performed on both sides, $A$ and $B$.}
\label{fig:circuit_purif}
\end{figure}
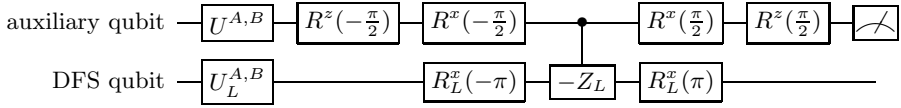
\end{center}
\end{widetext}
This circuit corresponds to the scheme proposed in~\cite{Deutsch96},
but here it is expressed in terms of operations that we can
implement in our setup. The circuit has to be applied on both ends
of the transmission line $A$ and $B$, where
\begin{equation}
U^A_L = \left({U^B_L}\right)^\dagger = R_L^x(\pi/2) \,.
\end{equation}
As in subsection~\ref{sec:R2}, the auxiliary qubit is again an atom
that controls a phase operation on the logical qubit. If the outcome
of the measurements at $A$ and $B$ coincide we end up with an
entangled DFS qubit pair of higher fidelity. If they do not coincide
the purification procedure has to be repeated.

\subsubsection{Module 4: Entanglement swapping between DFS qubits}
Entanglement swapping between neighboring DFS qubit pairs can be
done by a Bell measurement and subsequent local operations
$R_L^x(\pi/2)$ and $R_L^z(\pi/2)$ that are controlled by the outcome
of the measurement~\cite{Duer99}. The Bell measurement can be
decomposed in a controlled not (CNOT) gate, a Hadamard gate, and a
measurement in the computational basis (the Hadamard gate can be
replaced by, e.g., the sequence
$R_L^z(\pi/2)R_L^x(\pi/2)R_L^z(\pi/2)$).
\begin{widetext}
\begin{center}
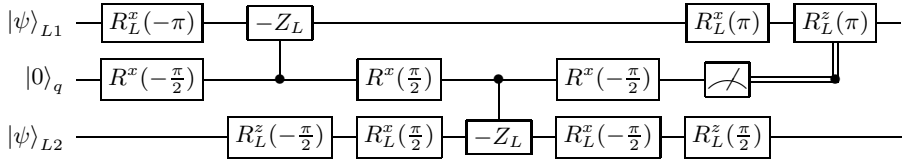
\begin{figure}[h!]
\[
\Qcircuit @C=1em @R=.7em
{
 & \lstick{\ket{\psi}_{L1}} & \gate{R_L^x(-\pi)} & \gate{-Z_L}  & \qw   & \qw         & \qw   & \gate{R_L^x(\pi)}& \gate{R_L^z(\pi)} & \qw\\
 & \lstick{\ket{0}_q}       & \Rxm               & \ctrl{-1}    & \Rxp  & \ctrl{1}    & \Rxm  & \meter           & \control \cw \cwx         \\
 & \lstick{\ket{\psi}_{L2}} & \qw                & \RzmL        & \RxpL & \gate{-Z_L} & \RxmL & \RzpL            & \qw               & \qw
}
\]
\caption{Quantum circuit for a CNOT operation between two logical qubits.}
\label{fig:circuit2}
\end{figure}
\end{center}
\end{widetext}
The implementation of a decoherence free CNOT operation is
relatively complicated including lengthy sequences of two-particle
interactions with pairwise individual tuning~\cite{Bacon00}.
Therefore we use an ancilla atom (as in Sec.~\ref{sec:R2}
and~\ref{sec:R3}) to mediate the interaction between different DFS
qubits. A possible quantum circuit for a CNOT gate between DFS
qubits is shown in Fig.~\ref{fig:circuit2}. The centerline in the
circuit represents the ancilla qubit. The shown circuit uses only
gates that we can implement in our scheme and leads to a CNOT gate
with the upper DFS qubit as the control qubit. We will discuss in
Sec.~\ref{sec:IIc} how the decoherence of the auxiliary atom
influences the fidelity of this operation.

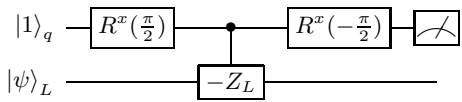
\begin{figure}[h!]
\[
\Qcircuit @C=1em @R=.7em {
  & \lstick{\ket{1}_{q}}  & \Rxp & \ctrl{1}   & \Rxm & \meter     \\
  & \lstick{\ket{\psi}_L} & \qw & \gate{-Z_L} & \qw & \qw & }
\]
\caption{Measurement of the logical qubit.}
\label{fig:circuit3}
\end{figure}

For the measurement of the DFS qubit we use the same ancilla as for
the CNOT gate. The corresponding circuit is shown in
Fig.~\ref{fig:circuit3}.
The final measurement of the ancilla is equivalent to a measurement of
the DFS qubit.

\subsubsection{Modules 5 and 6: State transfer between DFS qubits and
entanglement purification using pairs of DFS qubits
\label{Sec:Mod5u5}}
 The
state transfer of one DFS qubit to another can be done analogously
to Fig.~\ref{fig:circuit1a}, where $\ket{\psi}_q$ has to be replaced
by $\ket{\psi}_L$ . The necessary CPHASE gate between DFS qubits can
be constructed similarly as the CNOT gate: We merely have to omit
the single qubit rotations in the last line of
Fig.~\ref{fig:circuit2}. Also the entanglement purification scheme
of Fig.~\ref{fig:circuit_purif} can be adapted in this way. Again,
the auxiliary qubit is replaced by a DFS qubit and the CPHASE (or
CNOT) gate is realized as described above.

\subsection{Operation time and decoherence \label{sec:IIc}}
Some of the above operations require the use of a single auxiliary
qubit which is not decoherence free.
Entangling this qubit with the DFS qubits will therefore lower
the overall fidelity of the corresponding gate operations on the DFS
qubits.
The auxiliary qubit is represented
in our scheme by two long-lived internal states of an atom and thus
the major experimentally expected decoherence mechanism is
dephasing~\cite{Palma96}. This process is described by the master
equation
\begin{equation}
  \partial_t \varrho_\mathrm{aux} = - \frac{i}{\hbar} [\hat H,
  \varrho_\mathrm{aux} ] + \frac{\gamma}{2} \left( \sigma_z
  \varrho_\mathrm{aux} \sigma_z - \varrho_\mathrm{aux} \right)  \,,
\label{eq:master}
\end{equation}
where $\varrho_\mathrm{aux}$ is the density matrix of the auxiliary
atom, $\hat H$ is the Hamiltonian of the system, $1/\gamma$
characterizes the decoherence time, and $\sigma_z$ is a Pauli
matrix. Decoherence of the DFS qubits is not taken into account
here.

A quantitative measure of the operation fidelity is derived from
Ref.~\cite{Nielsen}. A distance measure of two density matrices
$\varrho$, $\sigma$ is given by
\begin{equation}
  f(\varrho, \sigma) = \left(\mathrm{Tr}\sqrt{\sqrt{\sigma} \varrho
  \sqrt{\sigma}}\right)^2 \,,
\end{equation}
where, in contrast to~\cite{Nielsen}, we use the squared trace since
for pure states this corresponds to a probability, i.e. the absolute
square of the overlap of the two states.  The operation- (or gate-)
fidelity can be derived from this quantity: Let $\mathcal{E}$ be a
quantum operation describing a quantum circuit where decoherence is
taken into account, and $\mathcal{E}_0$ the same operation without
decoherence. We define the operation fidelity by
\begin{equation}
  F(\mathcal{E}_0,\mathcal{E}) = \min_\varrho f(\mathcal{E}_0(\varrho),
  \mathcal{E}(\varrho))\,,
\end{equation}
where it is sufficient to minimize over pure input
states~$\varrho$~\cite{Nielsen}.

Let us first consider the state transfer from the atom to the DFS
qubit (Fig.~\ref{fig:circuit1a}).  To minimize decoherence effects,
the $R^x_L(\pi/2)$ and $R^x_L(-\pi/2)$ gates are applied before the
auxiliary atom interacts with the DFS qubit and after the
measurement, respectively. In addition, we neglect decoherence
during the $R^x(-\pi/2)$ rotation since the time to perform such a
gate on one atom is, in general, much smaller than the expected
decoherence time $1/\gamma$ (see Sec.~\ref{Sec:IIId}). Thus,
decoherence occurs mainly during the CPHASE gate. We examined the
dynamics of the system during the application of the CPHASE gate by
solving Eq.~(\ref{eq:master}) analytically. The resulting gate
fidelity for the CPHASE operation is given by
\begin{equation} \label{eq:fidel_CPHASE}
  F_{\text{CPHASE}}(\mathcal{E}_0, \mathcal{E}) = \frac{1 + \eh^{-\gamma
  t}}{2} \geq 1 - \frac{\gamma t}{2} \,.
\end{equation}
As will be discussed in Sec.~\ref{Sec:IIId} this behavior is
sufficient to ensure a working repeater scheme.

Note that, since unitary operations do not change the fidelity of
the circuits \cite{Nielsen}, all operations which only involve one
CPHASE gate have the same fidelity given in
Eq.~(\ref{eq:fidel_CPHASE}). This includes particularly the
operations shown in Figs.~\ref{fig:circuit_purif} and
\ref{fig:circuit3}.

In a similar way we calculated the fidelity of the CNOT gate between
the two DFS qubits (Fig.~\ref{fig:circuit2}). Again, all $R^{(x,z)}_L$
gates are applied before the auxiliary atom interacts with the DFS
qubit and after the measurement, respectively. During these periods
the auxiliary atom is not entangled with either DFS qubit and thus its
decoherence has no influence on them. Furthermore, we assume
again that decoherence during the $R^{(x,z)}$ rotations can be
neglected.  The gate fidelity of the CNOT operation is then given by
\begin{equation}
  F_{\text{CNOT}}(\mathcal{E}_0,\mathcal{E}) = \left( \frac{1 + \eh^{-\gamma t}}{2}
   \right)^2 \geq 1-\gamma t
  \,,
\end{equation}
where $t$ is the time which is needed to perform \emph{one} CPHASE
gate.  Since the two DFS qubits are not entangled before the CNOT
gate is applied it is sufficient to use only pure, unentangled
states for the minimization~\cite{Nielsen}. The fidelity for a
CPHASE gate between two DFS qubits is the same as for the CNOT gate
because they differ only by unitary operations.  As in the case of
the CPHASE gate, the decay of the CNOT fidelity is slow enough to
ensure a working quantum repeater (see Sec.~\ref{Sec:IIId}).

The fidelity of the state transfer between two DFS qubits described
in Sec.~\ref{Sec:Mod5u5} can also be calculated by explicitly
solving the time evolution. The state transfer fidelity
$F_\mathrm{ST}$ is given by
\begin{equation}
  F_\mathrm{ST}(\mathcal{E}_0, \mathcal{E}) = \frac{1}{4} \left(1+ \eh^{- 2\gamma t} \right)
  \left(1+ \eh^{- \gamma t} \right) \geq 1 - \frac{3}{2} \gamma t
  \,.
\label{ST}
\end{equation}

The fidelity of the entanglement purification using pairs of DFS qubits (see Sec.~\ref{Sec:Mod5u5})
is calculated using the chaining property for fidelity measures \cite{Nielsen}.
Assuming that the two DFS qubits on one side $A$ or $B$
are not entangled with each other we find as a lower bound
\begin{equation}
  F_\mathrm{EP}(\mathcal{E}_0, \mathcal{E}) \geq 1 - \left(
  \frac{3}{2} + \sqrt{2} \right) \gamma t \,.
\label{EP}
\end{equation}
In both formulas, Eqs.~(\ref{ST}) and (\ref{EP}), the time $t$ is
again the time needed for performing \emph{one} CPHASE gate.  As we
will show in Sec~\ref{Sec:IIId} all of these infidelities are small
for typical experimental decoherence times.

In summary, we have shown in this section that the DFS quantum
repeater can be divided into simple modules which require (apart
from single qubit rotations and measurements of an ancilla atom
which can be realized by standard methods) merely the operations
$R^x_L(\pm\frac{\pi}{2})$, $R^x_L(\pm\pi)$,
$R^z_L(\pm\frac{\pi}{2})$, $R^z_L(\pi)$, and a CPHASE gate on a DFS
qubit with the ancilla atom as control qubit.  In the following
section we proceed by describing how these basic operations can be
implemented in systems of trapped neutral atoms.


%% file: Implementation_new.tex
\section{Physical implementation in arrays of dipole traps\label{Sec:III}}

\begin{figure}
\begin{center}
   \includegraphics[width=6cm]{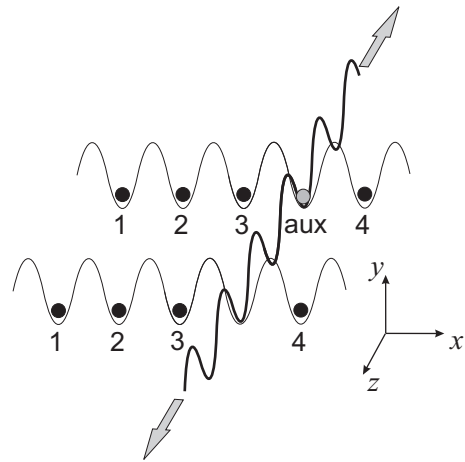} 
\end{center}
  \caption{Schematic picture of a repeater node. The four register atoms
  per qubit (black circles) are stored in a one-dimensional array of dipole
  traps, e.g., an optical lattice.
  Between the third and fourth atom there is a free lattice site in which
  the auxiliary atom (gray) can be moved, e.g., via an optical conveyor belt.
  \label{Fig_Repeater}}
\end{figure}

The physical system we consider for the implementation of a DFS
qubit is a one-dimensional chain of four atoms in an array of
optical dipole traps, e.g., an optical lattice \cite{Jaksch98} with
one atom per site. Each of the atoms has two long-lived internal
states $a$ and $b$ which encode a physical qubit $\hat a^\dagger
\ket{ \mathrm{vac}} \equiv \ket{0}$, $\hat b^\dagger \ket{
\mathrm{vac}} \equiv \ket{1}$.  In Fig.~\ref{Fig_Repeater} two DFS
qubits are schematically shown. The thick, diagonal line represents
an optical conveyor belt \cite{Kuhr01} storing the auxiliary atom.
This atom can be transported to the DFS strings where it occupies an
empty site between the third and the fourth atom, which is necessary
for the CPHASE operations described in Sec.~\ref{Sec:II}. Aside from
this, the auxiliary atom is needed for the state transfer of the
flying qubit to the DFS qubit: The flying qubits in our repeater
scheme are represented by entangled photon pairs. After a photon
arrives at a repeater node its state is transferred to the auxiliary
atom, e.g., by a cavity-enhanced
interaction~\cite{Cirac97,vanEnk-PRL-1997}. The atom is then moved
to the DFS qubit and the operation shown in Fig.~\ref{fig:circuit1a}
is performed. In order to implement the rotations on the DFS qubits
we also have to be able to selectively lower lattice barriers
between certain lattice sites. This can be done by using suitable
superlattices or by applying single laser pulses as described in
\cite{Pachos03}. The addressability of single atoms is only required
for the initialization of the register, cf. Sec.~\ref{Sec:IIIa}.

%
%

The general Hamiltonian for a one-dimensional atom chain for $M$
lattice sites (without auxiliary atom) is given by \cite{Jaksch98}
\begin{equation}
\begin{split} \label{Hamiltonian}
  \hat H =& \sum_{j=1}^{M-1} \left( -J_j^{(a)}\hat a_j^\dagger \hat a_{j+1}
  -J_j^{(b)}\hat b_j^\dagger \hat b_{j+1} + \mathrm{h.c.} \right) \\
  &+ \sum_{j=1}^{M}\left( \frac{U_a}{2} \hat a^\dagger_j \hat a^\dagger_j \hat a_j \hat
  a_j + \frac{U_b}{2} \hat b^\dagger_j \hat b^\dagger_j \hat b_j \hat
  b_j + U_{ab} \hat a_j^\dagger \hat a_j \hat b^\dagger_j \hat b_j \right),
\end{split}
\end{equation}
where $J_j^{(\alpha)}$ are state dependent hopping matrix elements
from site $j$ to site $j+1$ and $U_{a,b}$, $U_{ab}$ are state
dependent on-site interaction strengths. We should mention that for
the rotations $R^x_L$ and $R^z_L$ and for the CPHASE gate the
hopping constants for both kinds of atoms are required to be the
same, $J_j^{(a)} = J_j^{(b)}$, as well as $U_a = U_b = U_{ab}$, cf.
also Appendix~\ref{AppA}. Apart from the initialization of the
register we need $M=5$ lattice sites.

In the following subsections we will show how the register can be
prepared in state $\ket{0}_L$ and how the rotations $R_L^{(x,z)}$ and
the CPHASE gate can be implemented. These three ingredients are
sufficient to realize the repeater protocol.

\subsection{Initialization of state \protect{$\ket{0}_L$}\label{Sec:IIIa}}

The preparation of state $\ket{0}_\LL$ requires the creation of a
pair of singlet states $(\ket{01} - \ket{10})/\sqrt2$. These states
are decoupled from global operations but can be realized by local
operations and by selectively lowering the potential barriers
between neighboring atoms. Starting out from the product state
$\ket{0110}$ the potential barrier between the first and second atom
and between the third and fourth atom are lowered (e.g.~via a
superlattice potential). This generates two double well type
potentials. The parameters of the Hamiltonian
Eq.~(\ref{Hamiltonian}) are then given by
\begin{equation}
J_2 = 0, \, J_{1,3}^{(a)}=J_{1,3}^{(b)}=J,\,U_a=U_b=U_{ab}=U \,.
\label{cond1}
\end{equation}
Assuming that $J/U$ is sufficiently small (see also
Sec.~\ref{Sec:IIIb}), states with two atoms in one site can be
adiabatically eliminated leading to an effective Hamiltonian
\cite{duan03} of the form
\begin{align}
H_{\text{ini}} = -\frac{J^2}{U} ( &\sigma_1^x\sigma_2^x +
\sigma_1^y\sigma_2^y  + \sigma_1^z\sigma_2^z \nonumber\\
&\qquad+\sigma_3^x\sigma_4^x + \sigma_3^y\sigma_4^y  +
\sigma_3^z\sigma_4^z ).
\end{align}
Turning this Hamiltonian on for a time $t = \pi \hbar U / 8 J^2$ we
obtain the state $(\ket{01} +\ii\ket{10})\otimes(\ket{10} + \ii
\ket{01})/2$. Finally, a local operation on the first and fourth
physical qubit of the form $\exp(-\ii\pi\sigma_1^z/4)$ and
$\exp(-\ii\pi\sigma_4^z/4)$, respectively, yields the desired
product of singlet states. Alternatively, methods proposed in
\cite{Dorner03} can be used to realize the state $(\ket{01}
+\ket{10})\otimes(\ket{10} + \ket{01})/2$. Local single qubit
rotations of the first and fourth qubit can be used to bring this
state into the desired from. If laser addressing of single physical
qubits in adjacent lattice sites is not possible we can move the
first and the fourth atom into their neighboring, free sites (and
back again) by lowering the corresponding potential barriers.

In order to write information into the register we require a free
lattice site between the third and fourth atom. Hence, in a last
step the fourth atom is moved into the fifth lattice site by
lowering the potential barrier between the fourth and fifth site.


\subsection{Rotations of the logical qubit \label{Sec:IIIb}}

In this section we analytically investigate how to implement the
rotations $R^x_L$ and $R^z_L$ in the two limiting cases of a very
high interaction between the register atoms and a vanishing
interaction. The system is simulated numerically to check our
results.

\subsubsection{Analytical treatment}

The Hamiltonian $H_Z$ is, apart from a sign, equivalent to a swap of
the first two atoms $V_{12}$ [see Eq.~(\ref{eq:HZ})]. If we assume
high potential barriers between the second, the third, and the
fourth atom, i.e., $J_2^{(a,b)} = J_3^{(a,b)} =0$, such that these
atoms form a separate, decoupled system, we merely have to describe
the dynamics of the first and the second atom.  If we further assume
that $U_{a,b}, U_{ab}$ are much larger than the hopping terms
$J^{(a,b)}_j$ we can adiabatically eliminate all states with more
than one atom in a lattice site.  By choosing the parameters
according to
\begin{gather}
 J_1^{(a)}=J_1^{(b)}=J, \label{Cond_Z_J}\\
 U_a=U_b=U_{ab}=U,  \label{Cond_Z_U}
\end{gather}
we arrive at a the effective Hamiltonian
\begin{equation}
\begin{split}
H_Z &= -\frac{J^2}{U}
\left( \sigma_1^x\sigma_2^x + \sigma_1^y\sigma_2^y  + \sigma_1^z\sigma_2^z + \openone_{12} \right) \\
&= -\frac{2J^2}{U} V_{12} =  \frac{2J^2}{U} Z_L \,,
\end{split}
\end{equation}
where we omitted a constant term. The time evolution of the system
then generates the rotation
\begin{equation}
  R^z_L (\theta) = \exp \left({-\ii \frac{\theta}{2} Z_L}
  \right) \,,
\end{equation}
where $\theta = 4 J^2 t/U \hbar$. Hence, the time for a
$R_L^z(\pi)$-rotation is given by $t = \pi U \hbar / 4 J^2 $. As an
example we consider Na atoms in an optical lattice with a wavelength
of $\lambda = 514$nm. The height of the transverse potential
barriers are assumed to be $V_y = V_z = 50 E_\RR$, which is in the
range of today's experiments \cite{Folling-Nature-2005}. Here,
$E_\mathrm{R} = \hbar^2 (2 \pi)^2/2 m \lambda^2$ is the recoil
energy and $m$ is the mass of the atoms. If we increase the $s$-wave
scattering length by a factor of 5 using a Feshbach resonance and
choose $V_x = 7.7 E_\RR$ between the first and second lattice site
we get a ratio $U/J = 75$ which is well in the regime where the
adiabatic elimination is valid, as will be discussed in more detail
below. Note that for these parameters the assumptions for the
validity of the Bose-Hubbard Hamiltonian Eq.~(\ref{Hamiltonian}) are
fulfilled. The hopping term is then calculated to be $J=0.033E_\RR$
which results in an operation time of $t=8.7$ms for the
$R_L^z(\pi)$-rotation. We emphasize that we never leave the DFS
during the rotation and it is expected that this time is much
shorter than the decoherence time of the DFS qubit.

Nevertheless, the time for a $R_L^z(\pi)$-rotation can be made
considerably shorter by tuning the interaction between the register
atoms to zero (e.g., via a Feshbach resonance). As we show in
Appendix~\ref{AppB}, the qubits still remain in the DFS for all
times. In this interaction-free case, the exact dynamics of the two
atom system (i.e., without adiabatic elimination) results after a
time $t= \pi \hbar /2J$ in the mapping
\begin{equation}
  \ket{\alpha, \beta}_{12} \to - \ket{\beta, \alpha}_{12} \,, \,
  \alpha, \beta = 0,1 \,.
\end{equation}
Up to a global phase this is equivalent to a $R_L^z(\pi)$-rotation
on the logical qubits. The time needed for a lattice as above is $t
= 0.23$ms. The disadvantage of this scheme is that only rotations
with a phase of $\pi$ can be performed.

%
\begin{figure}
\begin{center}
  \includegraphics[width=6cm]{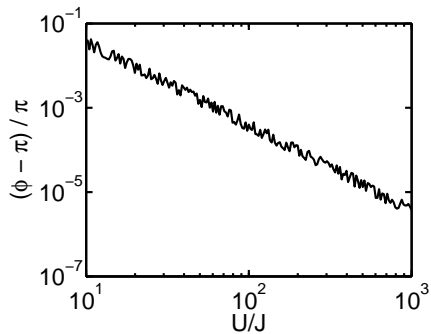}
  \end{center}
  \caption{Numerical studies of the phase difference
   $\phi$ between states $\ket{0}_\LL$ and $\ket{1}_\LL$
  after applying the rotation $R^z_\LL(\pi)$
  versus the ratio $U/J$ calculated with the full Hamiltonian
  Eq.~(\ref{Hamiltonian}) and the parameters
  Eqs.~(\ref{Cond_Z_J}) and (\ref{Cond_Z_U})
  given by adiabatic elimination. \label{Fig_Z_phase}}
\end{figure}
%

According to Eq.~(\ref{eq:HX}) the implementation of $H_X$ involves
three atoms. In the same spirit as for the $Z_L$-gate we require
that the potential barrier between the third and fourth lattice site
is high, i.e., $J^{(a,b)}_3 =0$. By choosing the remaining
parameters as
\begin{gather} \label{Cond_X_J}
  J^{(a)}_2=\sqrt{2} J^{(a)}_1=\sqrt{2} J^{(b)}_1 = J^{(b)}_2 \equiv
\sqrt{2} J \,, \\
  U_a = U_b = U_{ab} = U \,, \label{Cond_X_U}
\end{gather}
and requiring that $J/U\ll1$, we again adiabatically eliminate all
states with more than one atom in one well. A detailed calculation
is given in Appendix~\ref{AppA}. The resulting Hamiltonian on the
logical qubits is then given by
\begin{equation}\label{HamilX}
  H_X = - \frac{2 \sqrt{3} J^2}{U} X_L,
\end{equation}
where we omitted a constant term. This Hamiltonian
%
\begin{figure}
\begin{center}
  \begin{tabular}{ll}
    \qquad (a) & \qquad (b) \\
    \includegraphics[width=4cm]{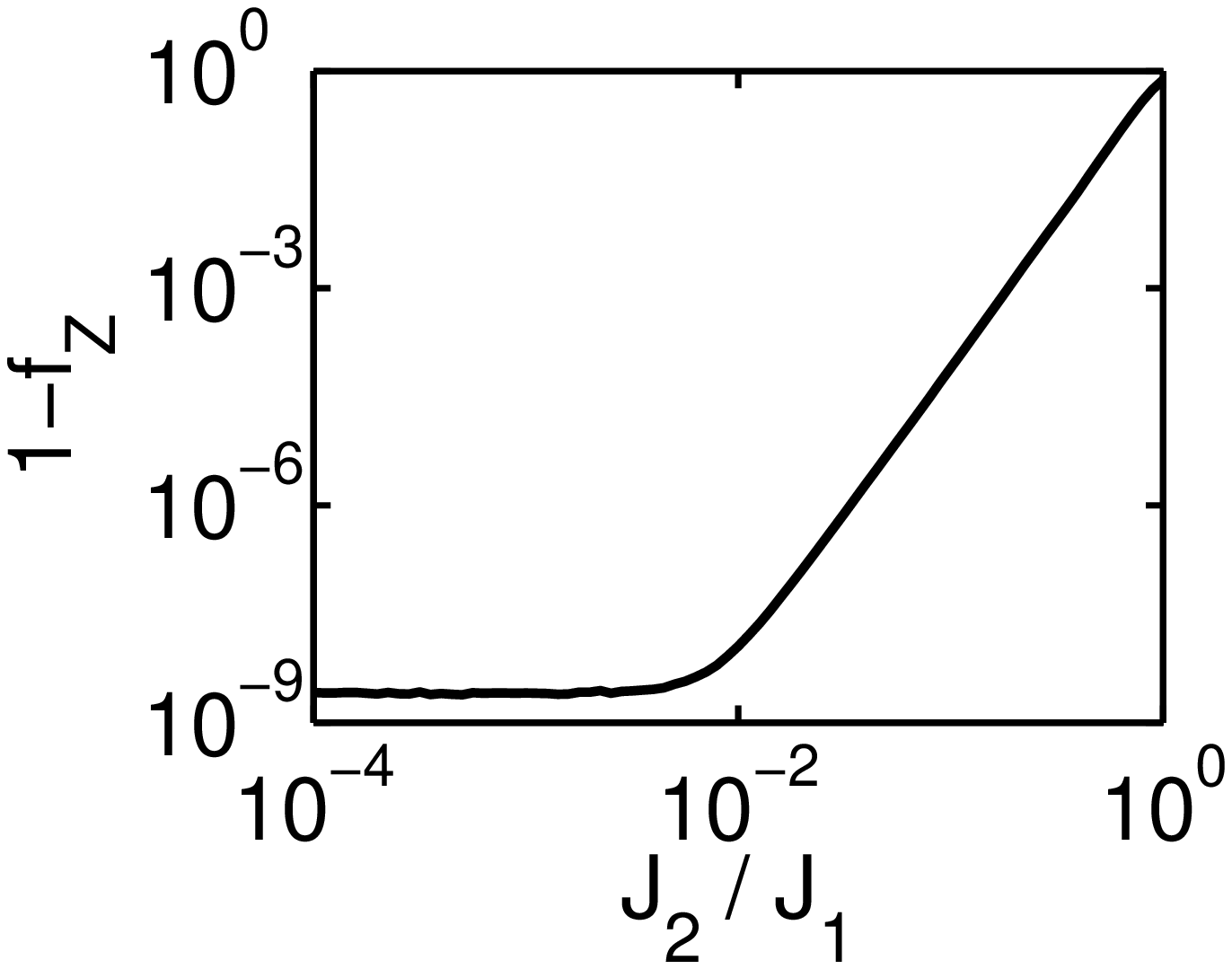} &
    \includegraphics[width=4cm]{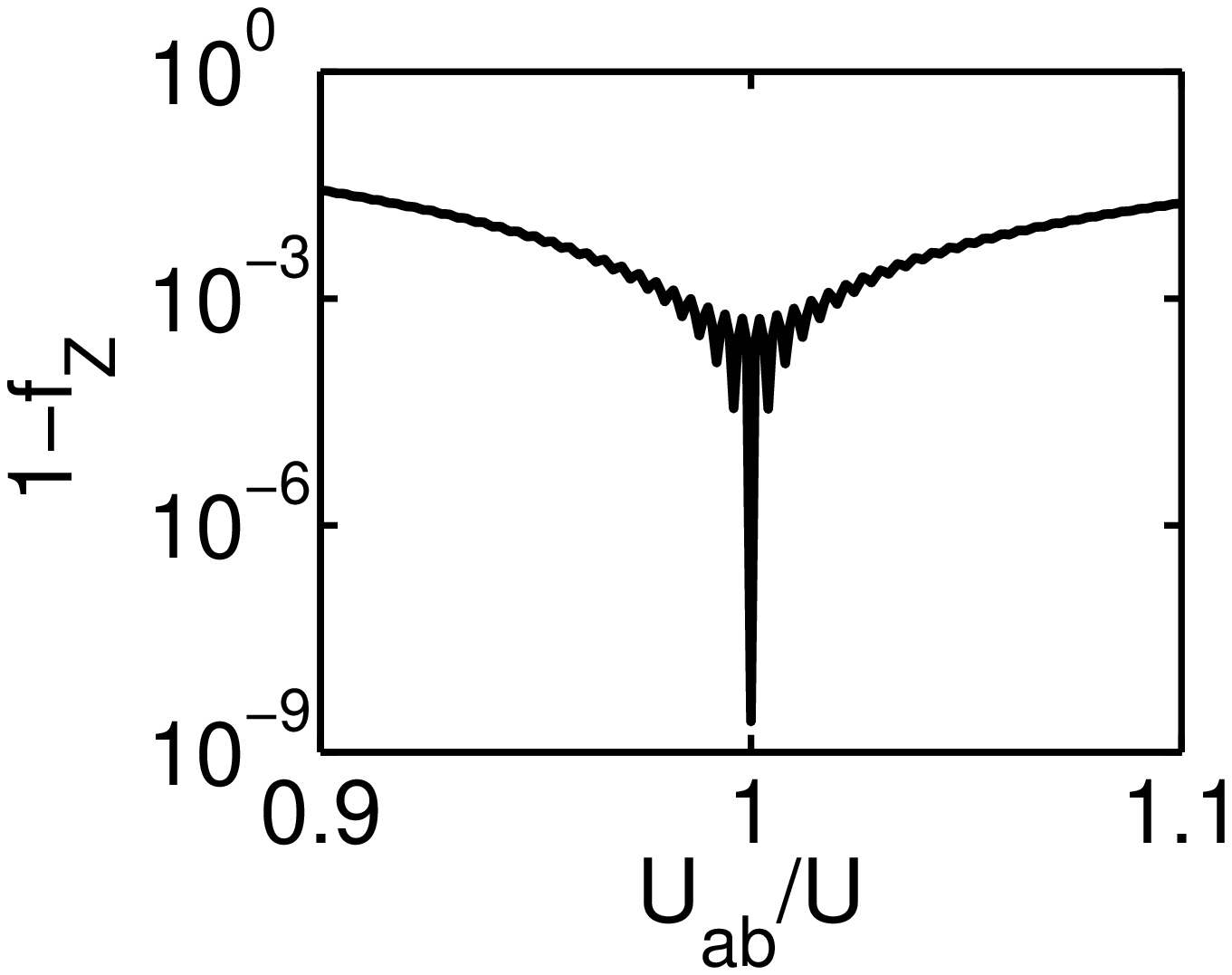}
  \end{tabular}
  \end{center}
  \caption{Fidelity Eq.~(\ref{fidelZ}) for the rotation $R^z_\LL(\pi)$.
  In (a) the hopping term between the second and third atom is
  detuned, in (b) the interaction between register atoms of kind $a$
  and $b$. The other parameters are (a) $U_{a,b} = U_{ab} = 100 J_1$,
  (b) $U_{a,b} = U = 100 J$.
\label{Fig_detune_Z}}
\end{figure}
%
generates a rotation
\begin{equation}
  R^x_L(\theta) = \exp\left(-\ii \frac{\theta}{2}
  X_L   \right) \,,
\end{equation}
where $\theta =- 4\sqrt{3} J^2 t/U \hbar$. Taking the numbers of the above example, the time required
for a $R_L^x(-\pi)$ rotation is $t = 5.0$ms.
\subsubsection{Numerical simulation}

To test the validity of the adiabatic elimination we simulated the
system numerically. In case of $R_L^z$-rotations the parameters of
the exact Hamiltonian Eq.~(\ref{Hamiltonian}) are chosen according
to Eqs.~(\ref{Cond_Z_J})-(\ref{Cond_Z_U}).We numerically calculated
the fidelity of the DFS qubit remaining in its state during an
$R_L^z$-rotation,
\begin{equation}\label{fidelZ}
  f_Z \equiv |\bra{l}_\LL \exp(-\ii H_Z t/\hbar) \ket{l}_\LL|^2
\end{equation}
with $l=0,1$. For $U/J = 75$ it turns out that the infidelity $1-f_Z$ is smaller than
$3\times10^{-3}$ for $l=1$ and smaller than $10^{-9}$ for $l=0$ during the whole duration of a $R_L^z(2\pi)$-rotation.

Since this fidelity does not take the phase difference $\phi$
between states $\ket{0}_L$ and $\ket{1}_L$ into account, $\phi$ was
calculated as well. In Fig.~\ref{Fig_Z_phase} the phase difference
is shown for the rotation $R^z_L(\pi)$.  As expected, for increasing
ratios $U/J$ the phase difference gets closer to $\pi$. For $U/J
\approx 75$ the inaccuracy of the achieved phase is already smaller
than $10^{-3}$, which shows that the adiabatic elimination is well
justified.
%
\begin{figure}
\begin{center}
  \includegraphics[width=6cm]{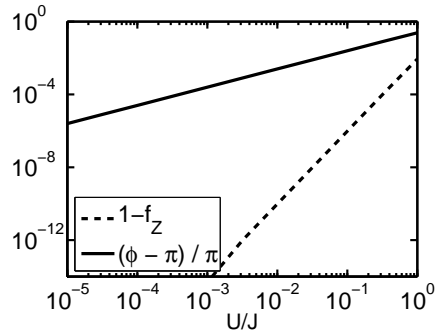}
  \end{center}
  \caption{Numerical studies for a fast rotation with no or only
  weak interaction between the atoms. The fidelity Eq.~(\ref{fidelZ}) of the
  rotation $R^z_\LL(\pi)$ and the phase difference between the two
  logical qubits after this rotation
  versus the ratio $U/J$ is calculated with the full Hamiltonian
  Eq.~(\ref{Hamiltonian}). The accuracy of the gate gets better with a smaller
  interaction $U$ between the atoms. Note that here no adiabatic elimination
  has been used. \label{Fig_Z0_fidelity}}
\end{figure}
%

From an experimental point of view it is also important to know what
happens when some of the parameters are detuned. Some results are
shown in Fig.~\ref{Fig_detune_Z}. The error due to the nonvanishing
hopping between the second and third lattice site $J_2$ is
negligible as long as $J_2 <10^{-2} J_1$. Because of this we have
completely neglected the influence of the fourth register atom which
is not involved in these rotations. In the case of a detuning of the
interaction coefficient $U_{ab}$ the fidelity changes quite quickly
if $U_{ab}/U$ differs from 1, but is smaller than $10^{-3}$ as long
as the detuning is less than 2\%.

The fidelity Eq.~(\ref{fidelZ}) and the phase difference $\phi$
between the logical qubit for the faster scheme of a
$R_L^z(\pi)$-rotation for small, residual interactions $U/J$ (in the
ideal case this interaction is vanishing) are shown in
Fig.~\ref{Fig_Z0_fidelity}. For this case the inaccuracy of the gate
decreases with decreasing $U/J$ and is smaller than $2\times
10^{-3}$ for $U/J \approx 10^{-2}$.  The error calculated according
to Eq.~(\ref{fidelZ}) is orders of magnitudes smaller than the phase
inaccuracy.

\begin{figure}
\begin{center}
  \includegraphics[width=6cm]{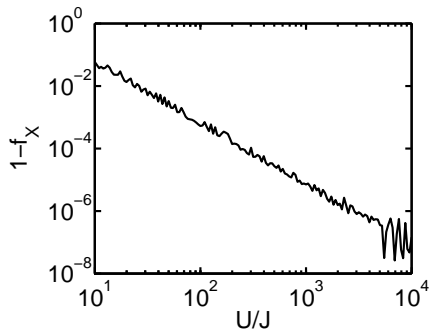}
  \end{center}
  \caption{Numerical studies of the fidelity Eq.~(\ref{fidelX}) of the
  rotation $R^x_\LL(\pi)$
  versus the ratio $U/J$ calculated with the full Hamiltonian
  Eq.~(\ref{Hamiltonian}) and the parameters
  Eqs.~(\ref{Cond_X_J}) and (\ref{Cond_X_U}) given by adiabatic
  elimination. \label{Fig_X_fidelity}}
\end{figure}

The simulations of the $R_L^x$-rotations lead to similar numbers.
The fidelity of this process, determined by
\begin{equation} \label{fidelX}
  f_X \equiv | \bra{0}_\LL  \exp(- \ii H_X t/\hbar) \ket{1}_\LL |^2,
\end{equation}
is shown in Fig.~\ref{Fig_X_fidelity} depending on the ratio $U/J$.
%
%
As can be seen from this figure, the infidelity $1-f_X$ decreases
with increasing ratio of $U/J$ and for a $U/J \approx 75$ the
infidelity is smaller than $10^{-3}$. Any infidelities because of a
change of the relative phases between the two logical states are
orders of magnitudes smaller and play no role.

\begin{figure}
\begin{center}
  \begin{tabular}{ll}
    \qquad (a) & \qquad (b) \\
    \includegraphics[width=4cm]{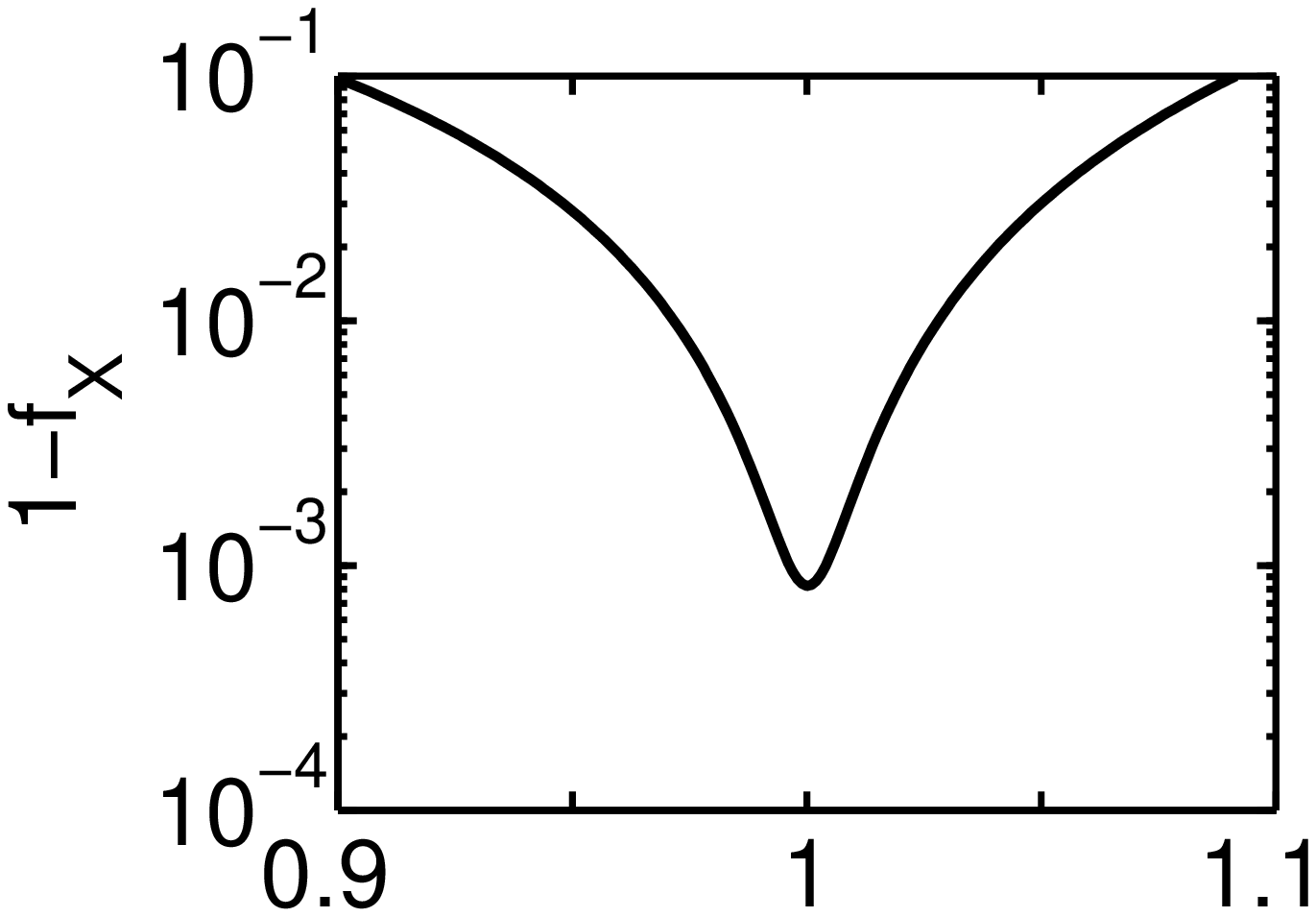} &
    \includegraphics[width=4cm]{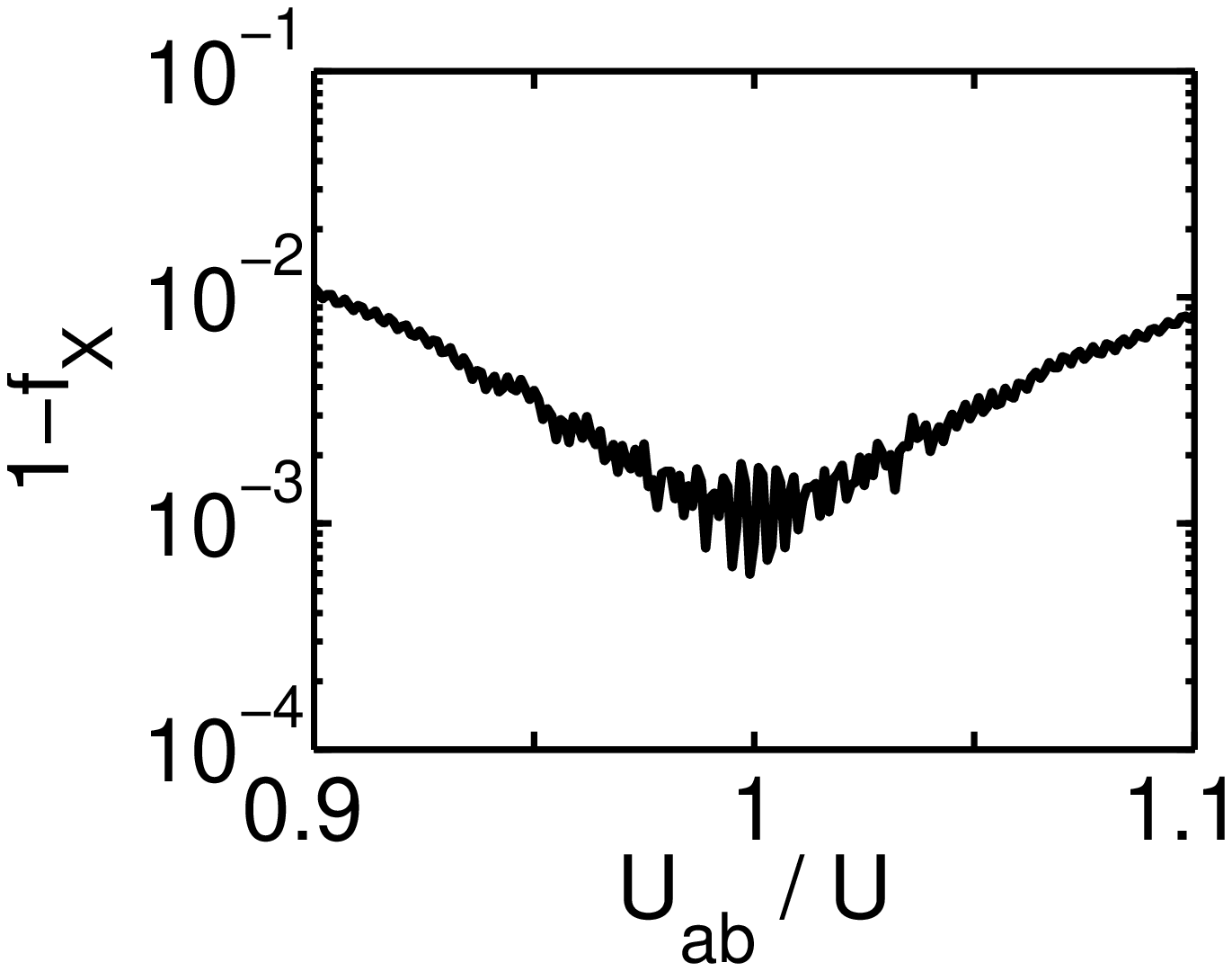} \\
    ~\qquad \hfill \raisebox{3.6ex}[-3.6ex]{$J_2 / \sqrt{2} J_1$}  \hfill~ & ~
  \end{tabular}
  \end{center}
  \caption{Fidelity Eq.~(\ref{fidelX}) for the rotation $R^x_\LL(\pi)$.
  In (a) the hopping term between the second and third atom is
  detuned, in (b) the interaction between register atoms of kind $a$
  and $b$. The other parameters are (a) $U_{a,b} = U_{ab} = 100 J_1$,
  (b) $U_{a,b} = U = 100 J$.
\label{Fig_detune_X}}
\end{figure}

Figure~\ref{Fig_detune_X} shows the results of numerical simulations
if various system parameters are detuned. For the case of a nonideal
hopping the deviation should be smaller than 1.5\% to yield
infidelities smaller than $3\times 10^{-3}$. A slight detuning in
the interaction strength is less critical, but should nevertheless
be smaller than 4\% to achieve small errors.

%

\subsection{Controlled phase gate \label{Sec:IIIc}}

The CPHASE gate can be realized by a controlled swap operation
between the third and the fourth atom since, as mentioned in
Sec.~\ref{SecII1}, $Z_L = -V_{34}$. We assume that the potential
barriers between the first, second, and third atom of the DFS qubit
are high enough to ignore the dynamics of the first two atoms. The
auxiliary atom is assumed to be distinguishable from the DFS atoms
and is confined in a sufficiently deep lattice such that it always
remains at its site. If the auxiliary atom is moved to the free site
of the DFS string (cf. Fig.~\ref{Fig_Repeater}) and if a register
atom tunnels into the site of the auxiliary they interact with each
other giving the Hamiltonian
\begin{equation}
  \hat H_\mathrm{CPHASE} = \hat H + \sum_{\sigma=0,1} U^q_\sigma
  \hat a^\dagger_4 \hat a_4
  \hat q^\dagger_\sigma \hat q_\sigma +U^q_\sigma \hat b^\dagger_4
  \hat b_4 \hat q^\dagger_\sigma \hat q_\sigma \,,
\label{Hamiltonian1}
\end{equation}
where $\hat H$ is the Hamiltonian given by Eq.~(\ref{Hamiltonian}),
$\hat q_\sigma$ is the state-dependent annihilation operator of the
auxiliary atom with $q$ its internal state, and $U^q_\sigma$ the
interaction strength which is independent of the internal state of
the DFS atoms. If the interaction $U^q_0$ for state $\ket{0}_q$ of
the auxiliary atom is very large compared to the hopping strength,
the third or fourth register atoms cannot enter the lattice site of
the auxiliary atom and hence they will remain in their sites.
Numerical tests show that during the gate operation this still holds
for an interaction strength of $U^q_0/J \approx 100 $, where $J$ is
the hopping term of the register atoms. Under these circumstances
the probability of finding the register in its original state is
larger than $1-10^{-3}$. If the auxiliary atom is in the state
$\ket{1}_q$ and $U^q_1 =0$ the existence of the auxiliary atom can
be ignored and we only need to solve the dynamics of two register
atoms in three lattice sites. A similar scheme has been discussed in
\cite{Micheli04,Daley-quant-ph-2005}, where an electromagnetically
induced transparency-like configuration was used to control the
interaction strength. We will investigate in the following the two
limiting cases of very large and vanishing interaction strength
between the register atoms.

\subsubsection{Large interaction between the register atoms}
For the case of large $U/J$ we calculate the fidelities of the swap
operation,
\begin{gather} \label{fidel01}
  f_{01} = \left|\bra{0\sigma1} \mathcal{U}(t) \ket{1\sigma0}\right|^2 \,, \\
  f_{00} = f_{11} = \left|\bra{0\sigma0} \mathcal{U}(t) \ket{0\sigma0}\right|^2 \label{fidel00}
\end{gather}
numerically, where $t$ is the numerically obtained time for the
CPHASE gate and $\mathcal{U}(t)= \exp(-\ii \hat H_{\mathrm{CPHASE}}
t/\hbar)$ is the time evolution operator corresponding to
Hamiltonian Eq.~(\ref{Hamiltonian1}). The states
$\ket{\alpha\sigma\beta},\,\alpha,\beta=0,1$ denote the states of
the third and the fourth DFS atom, $\ket{\alpha}$ and $\ket{\beta}$,
and $\ket{\sigma=1}$ is the internal state of the auxiliary atom,
located in between the two DFS atoms. We choose $J^{(a,b)} = J$ and
$U_{a,b} = U_{ab} = U$.

\begin{figure}
\begin{center}
  \begin{tabular}{ll}
    \qquad (a) & \qquad (b) \\
    \includegraphics[width=4.1cm]{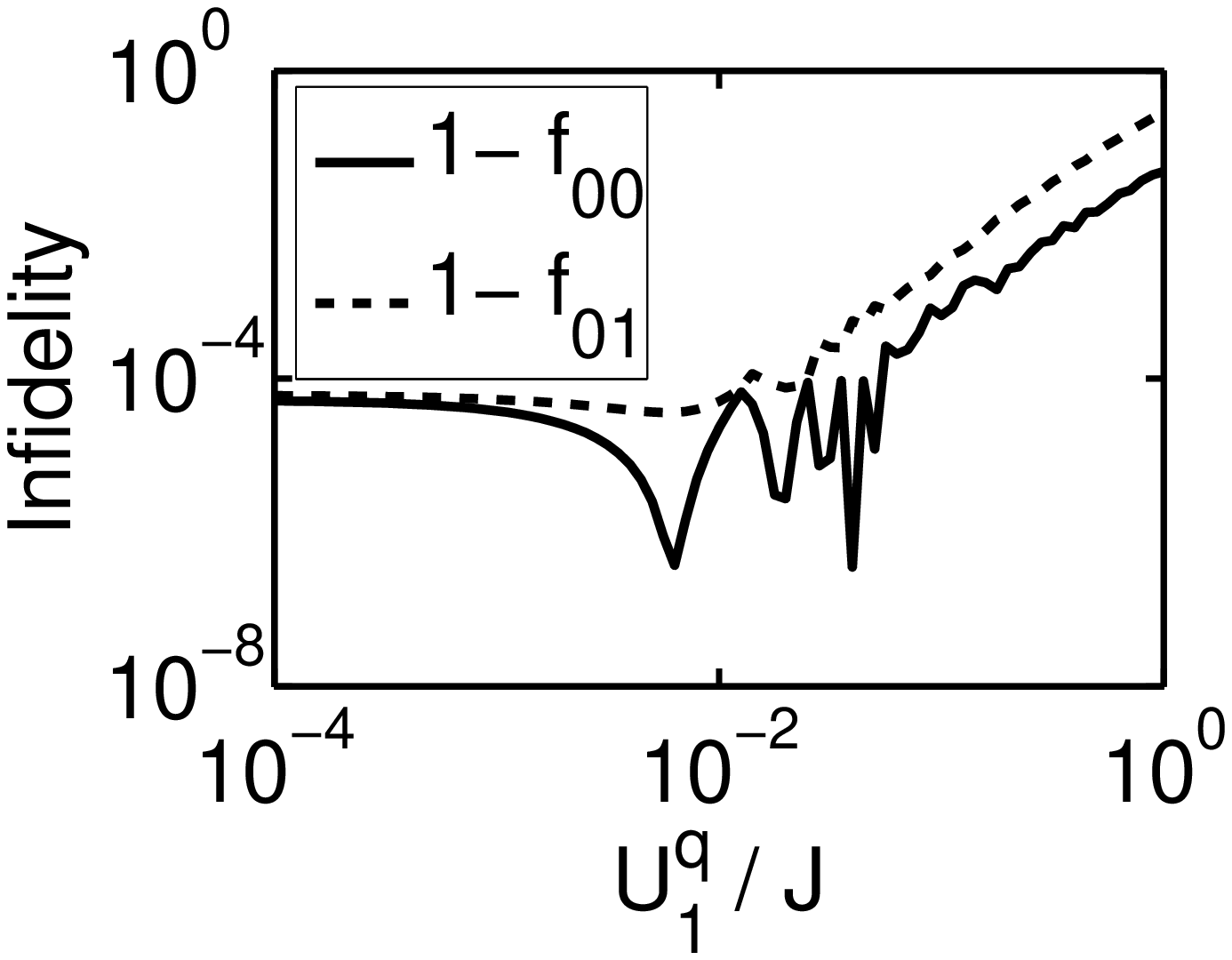} &
    \includegraphics[width=4.1cm]{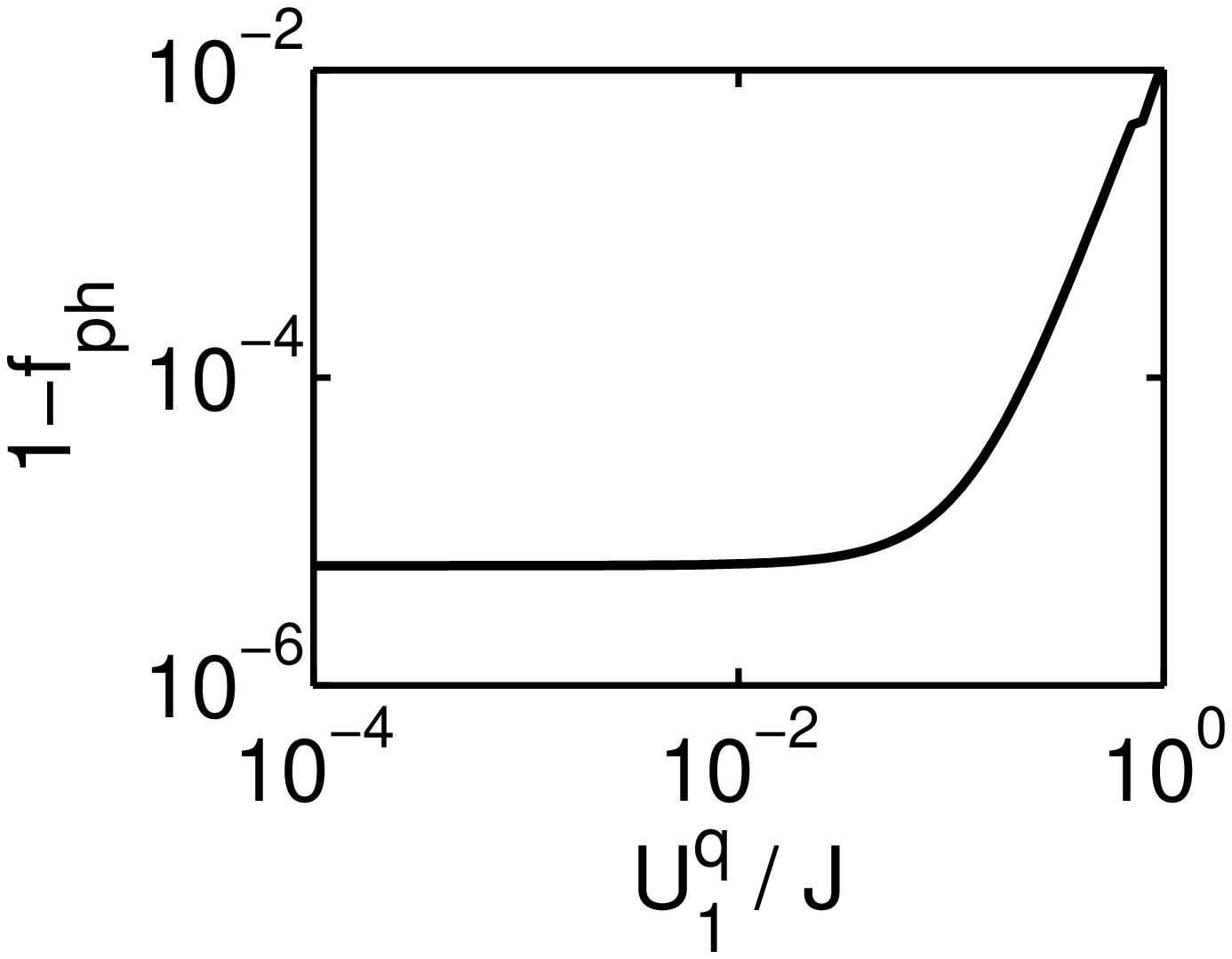} \\
  \end{tabular}
  \end{center}
  \caption{(a) Numerical results for the fidelities
  Eqs.~(\ref{fidel01}) and
  (\ref{fidel00}), where the
  auxiliary atom has a nonvanishing rest interaction with the register
  atoms.
  (b) Numerical results for the fidelity Eq.~(\ref{fidelph})
  due to the phase
  difference, where the
  auxiliary atom has a nonvanishing rest interaction with the register
  atoms. In both cases we took for the register atoms $U/J=100$.
   \label{Fig_marker_f}}
\end{figure}

In Fig.~\ref{Fig_marker_f}a the fidelities Eqs.~(\ref{fidel01})
and~(\ref{fidel00}) are shown for a small but nonvanishing
interaction $U^q_1$ between the auxiliary atom and the register
atoms for a ratio $U/J =100$. As can be seen the infidelities are
smaller than $10^{-3}$ as long as the interaction fulfills $U^q_1/J
< 0.05$. A further numerical observation is that the value of
$f_{00}$ ($f_{01}$) for very small $U^q_1/J$ gets smaller with
increasing $U/J$.
%
%

The fidelities Eqs.~(\ref{fidel01}) and (\ref{fidel00}) do not take
into account the occurrence of phases. In general, the dynamics of
the system will yield a phase,
\begin{equation}
  \ket{\alpha\sigma \beta} \rightarrow \eh^{\ii \varphi_{\alpha \beta}}
\ket{\beta \sigma \alpha}
\end{equation}
with $\alpha, \beta = 0,1 $. Since the interactions $U_\sigma^q$ do
not depend on the state of the DFS atoms we have $\varphi_{01} =
\varphi_{10} $ and $ \varphi_{00} = \varphi_{11}$ and a nonvanishing
$U_1^q$ leads to the behavior
\begin{align}
\mathcal{U}(t)\ket{0}_L &= \e^{\ii \varphi_{10}} \ket{0}_L \\
\mathcal{U}(t)\ket{1}_L &= \e^{\ii \varphi_{10}} \frac{1}{2\sqrt3}[ 2(\ket{1100} + \ket{0011}) \nonumber \\
& \qquad\qquad\qquad - \e^{\ii\varphi}(\ket{01}+\ket{10})^{\otimes
2}] \,,
\end{align}
of the four DFS atoms, i.e., to the presence of a nonzero phase
$\varphi \equiv \varphi_{00} - \varphi_{10}$. Thus the state
$\mathcal{U}(t)\ket{1}_L$ is not entirely contained in the DFS
anymore. To study this error we define the fidelity
\begin{equation} \label{fidelph}
  f_\mathrm{ph} = | \bra{1}_\LL \mathcal{U}(t) \ket{1}_\LL |^2 =
\frac{5 + 4 \cos\varphi}{9} \,,
\end{equation}
which is shown in Fig.~\ref{Fig_marker_f}b depending on $U_1^q/J$.
The infidelity $1-f_{\text{ph}}$ is typically one order of magnitude
smaller than the infidelities shown in Fig.~\ref{Fig_marker_f}a.

The time which is needed to perform the swap gate has been
calculated numerically. For a ratio of $U/J = 75$ and the same
lattice parameters as for the $R_L^z$-rotations we get $t = 11.4$ms.

Numerical tests show that for the same ratio $U/J=75$ the
probability of finding the qubit outside the DFS is smaller than $5
\times 10^{-4}$ during the whole gate operation time.

\subsubsection{No interaction between DFS atoms}

A substantially shorter gate time can be achieved for the case of a
vanishing interaction between the DFS atoms. In Appendix~\ref{AppB}
it is shown that the qubits still remain in the DFS during the gate
operation. By analytically solving the time evolution of the
Hamiltonian it can be shown (see also Appendix~\ref{AppB}) that
after a time $t=\pi\hbar/\sqrt{2} J$ the dynamics yields the mapping
\begin{gather}
  \ket{\alpha \sigma \beta} \to \ket{\beta \sigma \alpha} \,,\alpha,\beta = 0,1,\,\sigma =1
\end{gather}
which is a swap operation between the third and the fourth DFS atom
and thus a CPHASE gate. For the same lattice potential as earlier
with $V_x = 7.7 E_{\mathrm{R}}$ the gate time is given by $t =
0.33$ms.
%
\begin{figure}
\begin{center}
  \begin{tabular}{ll}
    \qquad (a) & \qquad (b) \\
    \includegraphics[width=4cm]{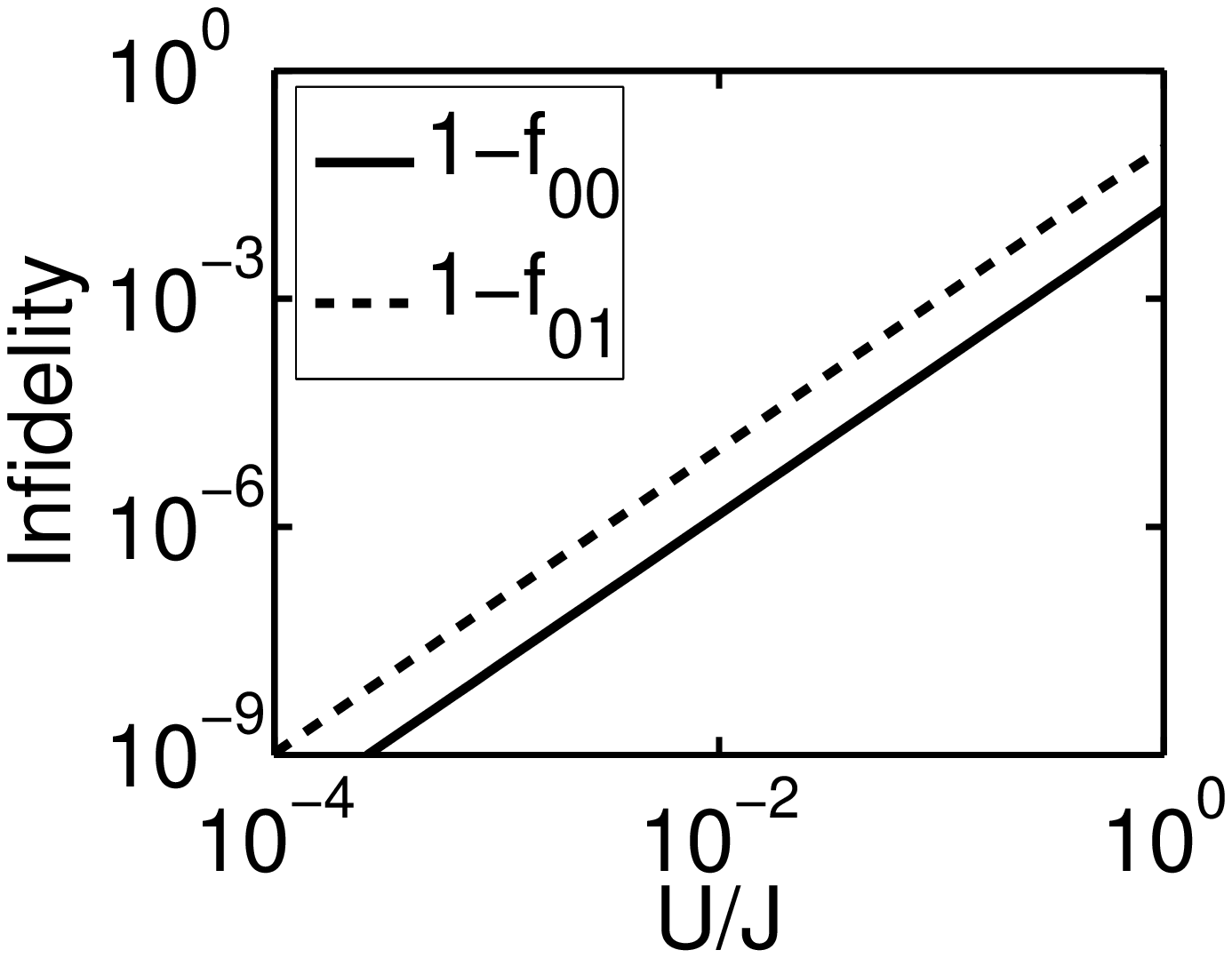} &
    \includegraphics[width=4cm]{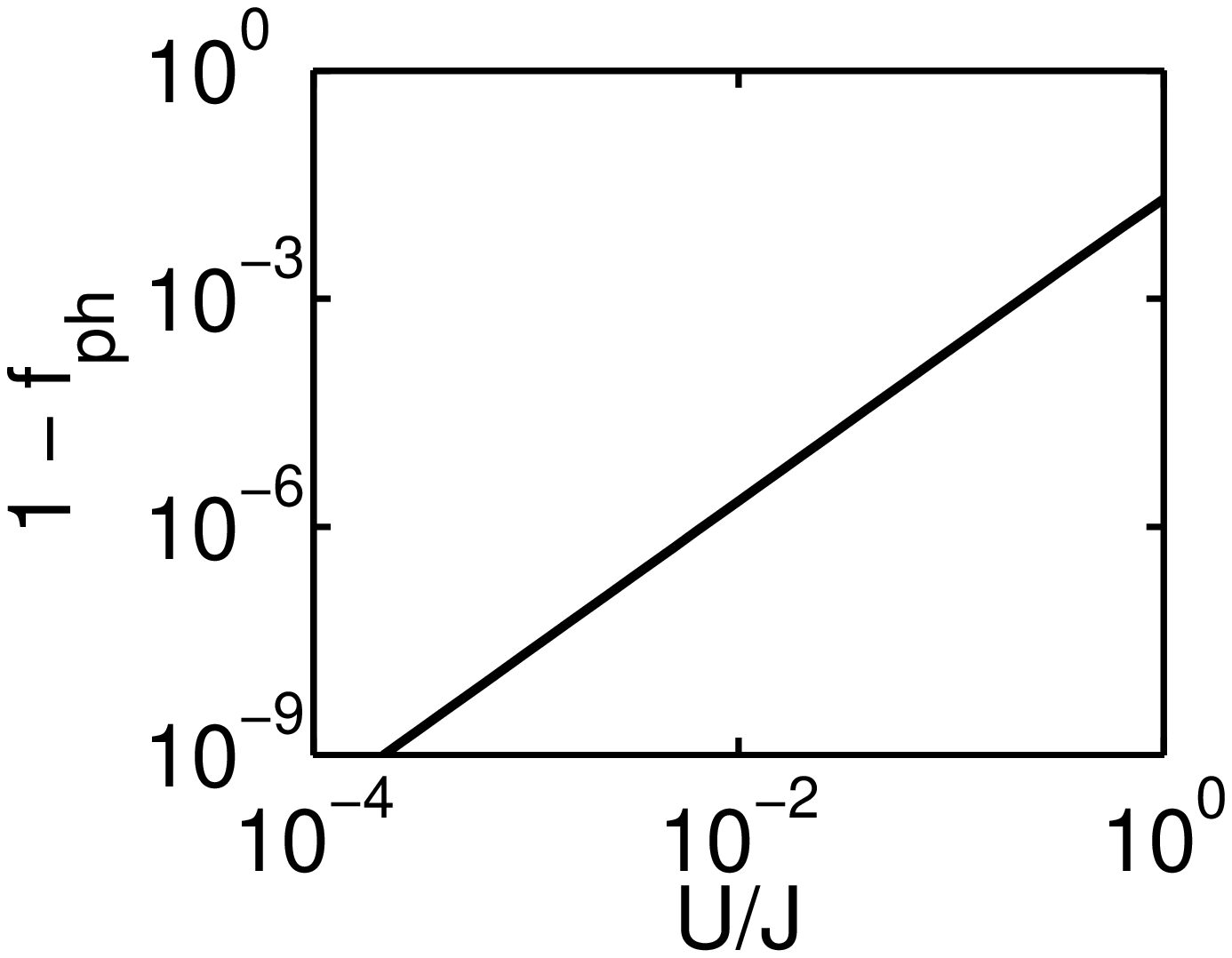}
  \end{tabular}
  \end{center}
  \caption{(a) Fidelity after evolving the state for a time $t=\pi
\hbar /\sqrt{2} J$ for the case of noninteracting register atoms.
(b) Fidelity Eq.~(\ref{fidelph}) of state $\ket{1}_\LL$ after the
time evolution, where a perfect fidelity of the atom swap operation
has been assumed and the error is only due to the phase difference,
see text. \label{Fig_marker_U0}}
\end{figure}
%

The results of numerical tests for a small rest interaction between
the register atoms are shown in Fig.~\ref{Fig_marker_U0}.
Figure~\ref{Fig_marker_U0}a shows the infidelities derived form
Eqs.~(\ref{fidel01}) and~(\ref{fidel00}) and
Fig.~\ref{Fig_marker_U0}b shows the infidelity derived from
Eq.~(\ref{fidelph}) all depending on $U/J$. The fidelities are
better than $1-10^{-3}$ for interactions $U/J < 0.05$. The results
of numerical simulations for a nonvanishing rest interaction between
the DFS atoms and a state $\ket{1}_q$ of the auxiliary atom yield an
infidelity smaller that $3\times 10^{-3}$ for $U^q_1/J\leq0.05$.

In conclusion, a reliable CPHASE gate can be implemented in both
cases.  If the auxiliary atom is in the state $\ket{1}_q$ and thus a
swap of register atoms 3 and 4 should occur, the interaction $U^q_1$
between the auxiliary atom and the register atoms is required to be
smaller than $0.05J$.  If the auxiliary atom is in the state
$\ket{0}_q$ no swap should occur. This holds during the gate
interaction time for an interaction $U^q_0$ larger than
approximately $100 J$. The difference of more than three orders of
magnitude in the interaction can be achieved either by using a
Feshbach resonance or by using an electromagnetically induced
transparency scheme as in \cite{Micheli04}.


\subsection{Gate times and decoherence \label{Sec:IIId}}

Recent experiments have demonstrated that the decoherence time of
quantum information stored in DFSs can be increased by several
orders of magnitudes compared to unprotected qubits.  In the case of
ion traps decoherence times of more than 7s have been reported
\cite{Langer05} and although the DFS scheme in this experiment was
different from ours we expect a similar improvement in the
reliability of our quantum memory. The required storage time is
found by adding the single qubit and CPHASE gate times calculated in
previous sections. According to these the state transfer from the
auxiliary atom to the DFS takes 11ms, neglecting the time which is
needed to implement the rotation and measurement of the auxiliary
atom. The entanglement purification module described in
Fig.~\ref{fig:circuit_purif} takes 26ms, the CNOT operation of
Fig.~\ref{fig:circuit2} takes 28ms. The necessary readout process of
the DFS qubits takes 11ms. In comparison to these times the
additional operations in steps (i), (ii), and (iii) of
Fig.~\ref{fig:repeater} are expected to contribute a negligible
amount of extra time to the overall scheme
\cite{Pan98,Jennewein02,Sciarrino02,Takei05,Gisin05,Cirac97,vanEnk-PRL-1997}.
All these times are very short compared to the expected decoherence
time of the DFS qubit. Since the system remains in the DFS during
the entire single qubit operations we conclude that decoherence of
the DFS qubits plays only a minor role.

However, in the modules which involve a CPHASE gate between the
auxiliary qubit and a DFS qubit the decoherence of the auxiliary
atom has to be taken into account since, in general, it gets
entangled with the DFS qubit thus lowering the overall fidelity.  As
discussed in Sec.~\ref{sec:IIc} the fidelity for, e.g.,  the CNOT
gate is better than $1-\gamma t$, where $\gamma$ is the inverse
decoherence time and $t$ is the time to perform a CPHASE gate.  In
our fastest example (without interaction between register atoms) a
CPHASE gate takes approximately 0.33ms. In Ref.~\cite{Kuhr05}
decoherence times of 146ms for $\sim 50$ Cs atoms stored in a
red-detuned standing wave have been reported. In the experiment
presented in Ref.~\cite{Kuhr03} the influence of a moving conveyor
belt is investigated. It is found that the transport procedure
decreases the coherence time by approximately a factor of 2.  This
would still be long enough for our purposes. For the entanglement
purification module, which has the worst fidelity of all our models,
the influence of the auxiliary atom leads to a fidelity of better
than 98.7\%. For the other modules the fidelity due to decoherence
of the auxiliary atom is better than 99\%. The usage of a
blue-detuned trap could give further considerable increase of the
decoherence time. Aside from this, it should be emphasized that all
these times have been measured for many atoms in a trap.  In the
case of single atoms longer decoherence times can be expected. Thus,
the influence of the auxiliary atom on the modules described in
Sec.~\ref{Sec:AllModules} is small and only slightly decreases their
fidelity.

%